\documentclass[letterpaper,english,aps, prb, eprint, longbibliography, floatfix, twocolumn, nobalancelastpage, flushbottom]{revtex4-2} % 2-column reprint, with HUB modifications
\usepackage{xcolor}
\usepackage{amsmath}
\usepackage{amssymb}
\usepackage{graphicx}

\usepackage[pdfusetitle, bookmarks=true, bookmarksnumbered=false, bookmarksopen=false, 
 breaklinks=true, pdfborder={0 0 0}, pdfborderstyle={}, backref=false, colorlinks=true, citecolor=blue]
 {hyperref}
\usepackage[all]{hypcap}% let hyperlinks correctly point to figures rather than captions; must be loaded after the hyperref package! 
\usepackage{orcidlink} %% must be loaded after hyperref and xcolor!

\usepackage{filecontents}     %% Included in Overleaf but not necessarily in TexShop or Other



\begin{document}
\title{Quantum {Hall} {Andreev} Conversion in Graphene Nanostructures}
\author{Alexey Bondarev}
\affiliation{Department of Physics, Duke University, Durham, North Carolina 27708-0305, USA}
\author{William H. Klein}
\affiliation{Department of Physics, Duke University, Durham, North Carolina 27708-0305, USA}
\author{Harold U. Baranger\,\orcidlink{0000-0002-1458-2756}}
\email{harold.baranger@duke.edu}
\affiliation{Department of Physics, Duke University, Durham, North Carolina 27708-0305, USA}

%% should be hard-wired to date of submission to arXiv
%\date{\today} %% should be hard-wired to date of submission to arXiv
\date{2 January 2026; \href{https://doi.org/10.1103/kp8y-y288}{Phys. Rev. B 113, 085417 (2026)} (accepted version)}

\begin{abstract}
We study Andreev conversion in clean nanostructures containing an interface between graphene in the quantum Hall (QH) state and a superconductor, focusing on the lowest Landau level.  First, several graphene nanostructures formed from zigzag edges with sharp corners are considered using a tight-binding model.  We find the scattering state for an electron impinging on the interface from the upstream QH edge state, together with the probability of it exiting as a hole in the downstream QH edge state (Andreev conversion).  From these results, we deduce the behavior for edges at an arbitrary angle and for rounded corners.  A key issue is whether the graphene-superconductor interface is fully transparent or only partially transparent.  For full transparency, we recover previous results.  In contrast, interfaces with partial but substantial transparency (well away from the tunneling limit) behave very differently: (i) the hybrid electron-hole interfacial modes are not valley degenerate and (ii) intervalley scattering can occur at the corners, even when rounded.  As a result, interference between the two hybrid modes can occur, even in the absence of disorder.  Finally, we compare the sensitivity of Andreev conversion to interface transparency in the QH regime to that in the absence of a magnetic field.  While the zero-field result closely follows the classic Blonder-Tinkham-Klapwijk relation, Andreev conversion in the QH regime is more robust. 
\end{abstract}

\maketitle

\section{Introduction}
\label{sec:Intro}

\subsection{Context and Motivation}

Nanostructures that incorporate a superconducting element, causing Andreev conversion from electrons to holes \cite{deGennesBook, vanWees_Sem-SupReview_chapter97, BeenakkervanHouten-SuperQPC92, PannetierCourtoisJLTP00, Klapwijk_ProximityAndreev_2004, Asano_AndreevBook21}, are attracting increasing attention for both devices \cite{WenShabani_LogicIEEE19, PhanShabani_PampPRAppl23, Huang-GraphDetect-X24} and basic physics \cite{VignaudSacepe_ChiralSuper_Nat23, LingfeiGlebLossPRL23, HatefipourShabani_Andreev-QH-QPC_PRB24}.  
The desire for maximum flexibility and control suggests investigating gateable nanostructures in which an applied potential changes the carrier density locally.  This then leads naturally to studying superconducting contacts to two-di\-mensional (2D) electron systems \cite{BeenakkervanHouten-SuperQPC92, vanWees_Sem-SupReview_chapter97, DohDeFranceschi_TunableIS_Sci05}: the two main experimental systems that have emerged are contacts to either InAs-based heterostructures \cite{HatefipourShabani_QH-S_NL22} or graphene \cite{LeeLeeReviewRPP18}.  

Well-defined, robust channels for incoming and outgoing particles are a significant advantage in interpreting and controlling properties of nanostructures.  In this regard, chiral quasi-one-dimensional modes are particularly attractive since there is no backscattering.  These can be created, for instance, by inducing the quantum Hall (QH) effect in the 2D electron system---the development of electron quantum optics is a striking example \cite{ElectronQOptReview14, SplettstoesserHaug_pssb17, Chakraborti-EQopt_JPCM24}.  
QH states can coexist with superconductivity when the magnetic field is less than the critical field of the superconductor.  Graphene is particularly attractive since the QH effect occurs at modest fields.  The incorporation of the QH effect adds interesting physics issues of its own, e.g., an interface between topological and trivial quantum states and a first step toward potential one-dimensional Majorana modes \cite{QiHuguesZhangPRB10, LianZhangChiralMajPNAS18}.  

Interest in graphene-superconductor interfaces developed shortly after the discovery of graphene \cite{BeenakkerRMP08, LeeLeeReviewRPP18}.  The development of good contacts from metals to the edge of graphene clad by BN protective layers \citep{WangDean_1DcontactSci13} led to increasing experimental activity \cite{CaladoDelftNatNano15, BenShalomManchesterNatPhy16, IvanGlebBallisticPRL16,  ParkLeePRL18}, including on QH graphene/superconductor interfaces \cite{FrancoisGlebSci16, ParkLee_S-QHedge_SciRep17, SahuBangalorePRL18, LeeHarvardNatPhy17, LingfeiGlebCAESNatPhys20}. 
Recent experiments in nanostructures have studied the downstream resistance \cite{LeeHarvardNatPhy17, LingfeiGlebCAESNatPhys20, SahuDas_Noise_PRB21, LingfeiGlebLandButPRB24}, which directly probes the interfacial Andreev conversion.

While theoretical work on the QH-superconductor interface predates graphene \cite{MaZyuzin_S-QH_EPL93, TakagakiQHEPRB98, HoppeZulickePRL00, GiazottoPRB05}, it expanded in response to the experimental possibilities \cite{AkhmerovValleyPolarPRL07, KhaymovichEPL10, StoneLinPRB11, VanOstaayTripletPRB11}.  
 In particular, recent experiments in both graphene and InAs heterostructures have stimulated further theoretical work on the identification of Andreev edge states in various transport probes 
 \citep{LianEdgestatePRB16, GamayunCheianovPRB17, Peralta-Bariloche_EdgeTran_PRB21, MichelsenSchmidt_SupercurEnable_PRR23, Khrapai_BogoliubovInterferom_PRB23, DavidGrenoble_Geometrical_PRB23,  CuozzoRossiPRB24}
 %, AlexeyB_InfInter_PRB25} 
 and on the role of non-idealities like disorder, dissipation, and decoherence on transport signals 
 \citep{Manesco-QHgrS_SP22, KurilovichDisorderAES_NCom23, TangAlicea_VortexEnabled_PRB22,  SchillerOreg_FermDissip_PRB23, HuLian-Decohere-X24}. 

Nanostructures with a QH graphene/superconductor interface were first studied theoretically for perfect Andreev reflection at an ideal interface \cite{AkhmerovValleyPolarPRL07}, which was treated using a valley-independent boundary condition \cite{TitovBnkr06} so that intervalley scattering is absent.  
The result for Andreev conversion---the probability, $P_{he}$, that an incoming electron exits as a hole---is 
\begin{equation}
	P_{he}=\left\{ \begin{array}{l}
		1, \textrm{same type of edge in and out}\\
		0,  \textrm{opposite edge types}
	\end{array} \right. ,
\label{eq:Phe-ideal-AkhBeen}
\end{equation} 
where ``type of edge'' refers to the effective zigzag type of a graphene edge meeting the interface (whether the majority of the terminal atoms are in the A or B sublattice \cite{AkhmerovValleyPolarPRL07, AkhBeenBoundaryPRB08}) and one compares the type of the edge for the incoming mode to that for the outgoing mode.  [Eq.\,(\ref{eq:Phe-ideal-AkhBeen}) does not apply to the rare case of an armchair edge.]  

More recent work raises questions about the usefulness of Eq.\,(\ref{eq:Phe-ideal-AkhBeen}) as a guide for non-ideal nanostructures.  For example, it has been recognized that sharp corners in semiconductor nanostructures strongly influence Andreev conversion \cite{DavidGrenoble_Geometrical_PRB23}.  Since such geometric effects clearly cannot appear in (\ref{eq:Phe-ideal-AkhBeen}), are they absent in graphene?  
As a second example, in our previous work on QH graphene-superconductor interfaces \cite{AlexeyB_InfInter_PRB25}, we showed that the propagating modes at non-ideal graphene-superconductor interfaces lack several special properties of ideal interface modes, in particular valley degenerate dispersion and equal weight of electrons and holes.  
%Does the absence of these special properties lead to modifications of Eq.\,(\ref{eq:Phe-ideal-AkhBeen})? 
How is Eq.\,(\ref{eq:Phe-ideal-AkhBeen}) modified when these special properties are absent? 
To clarify the relation between non-ideal interfaces, scattering at corners, and Andreev conversion, here we study QH graphene-superconductor nanostructures of different shapes with both ideal and reduced interface transparency.  

\begin{figure}
\includegraphics[width=2.0in]{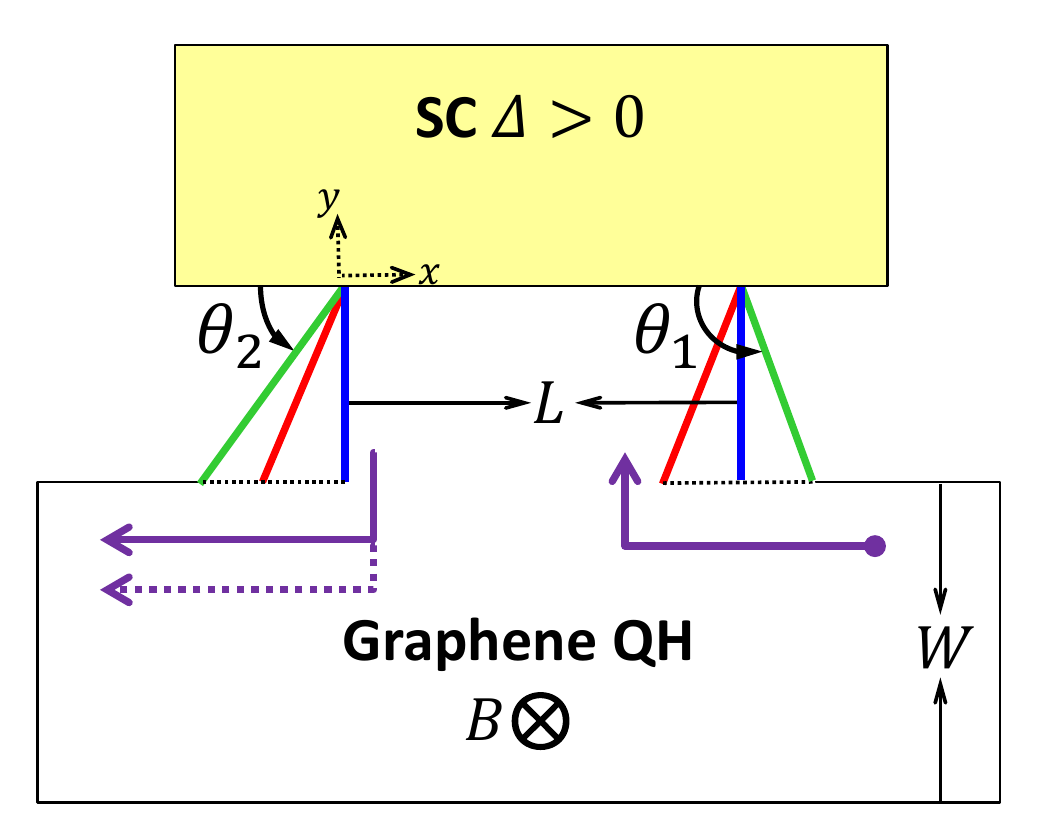}
\caption{Geometry of the nanostructure and interface.  Different nanostructures connecting the white graphene nanoribbon to the yellow superconductor are color coded: parallelogram (red), trapezoid (green), and rectangle (blue).  Angles $\theta_{1}$ and $\theta_{2}$ could but need not be identical.  (The $\theta_{1}\!=\!\theta_{2}\!=\! 60^\circ$ parallelogram with zigzag edges is our canonical case.)} 
\label{fig:Geometry}
\end{figure}

\subsection{Summary of Main Results}

In this work, we find the scattering state and Andreev conversion probability in the lowest Landau level (LLL) for several simple nanostructures in the geometry of Fig.\,\ref{fig:Geometry}.  The nanostructures have a single gra\-phene-superconductor interface and no disorder.  
We solve the Bogoliubov-de\,Gennes (BdG) equation \cite{deGennesBook, BeenakkerRMP08, Asano_AndreevBook21} for a tight-binding model using a honeycomb lattice for graphene and a square lattice (near half filling) for the superconductor, finding the scattering state wavefunction and S-matrix.  

\begin{figure}
\includegraphics[width=2.5in]{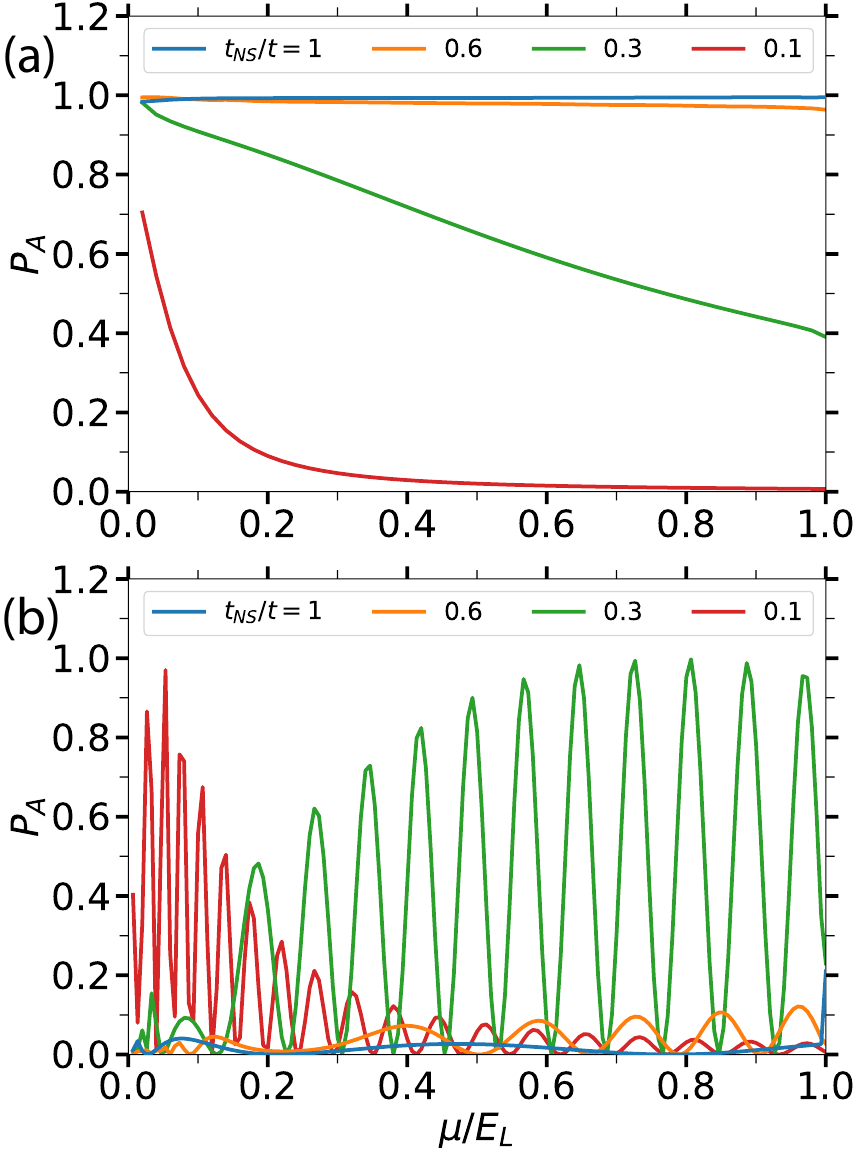}
\caption{Andreev conversion probability, $P_{he}$, as a function of the graphene edge-state filling $\mu_\textrm{gr}/E_{L}$ for two zigzag nanostructures: (a) parallelogram and (b) trapezoid, corresponding respectively to red and green lines in Fig.\,\ref{fig:Geometry}.  The strength of graphene-superconductor coupling, $t_{NS}/t$, is indicated by the line color.  At full interface coupling, $P_{he}(\mu_\textrm{gr})$ is constant, either $0$ or $1$.  At partial coupling, intervalley-scattering occurs at some corners.  Interference oscillations are present when the corners at the entrance and exit of the interface \emph{both} cause intervalley scattering, as in the trapezoid.  
($E_L\!\equiv\!\sqrt{2}\hbar v/l_{B}$; standard parameters Sec.\,\ref{subsec:Parameters}. Further discussion of each case is in Secs.\,\ref{subsec:Parallelogram} and \ref{subsec:Trapezoid}, respectively.)
}
\label{fig:SummaryPA} 
\end{figure}

The Andreev conversion probability as a function of the graphene chemical potential, $P_{he}(\mu_\textrm{gr})$, provides a useful way to summarize the results.  Fig.\,\ref{fig:SummaryPA} shows results for two nanostructures at four values of the graphene-superconductor interfacial coupling ($t_{NS}/t$). 

For a well-matched, transparent interface ($t_{NS}/t\!=\!1$), we find that the scattering wavefunction is always smooth at both the incoming and outgoing corner of the interface.  This gives rise to either complete or zero Andreev conversion for all $\mu_\textrm{gr}$ (blue lines), thus verifying Eq.\,(\ref{eq:Phe-ideal-AkhBeen}).  

In contrast, for an interface with reduced transparency, intervalley scattering may occur at the corners on either end of the interface.  Scattering by such corners is evident in the scattering state and influences Andreev conversion substantially.  If only one such corner lies at the interface, $P_{he}(\mu_\textrm{gr})$ is reduced and varies monotonically, as in Fig.\,\ref{fig:SummaryPA}(a).  However, if \emph{both} interfacial corners cause intervalley scattering, interference between the two interfacial modes occurs, and $P_{he}(\mu_\textrm{gr})$ oscillates rapidly [panel (b)]. 
We show how intervalley scattering enables Andreev conversion despite reduced interface pairing.  This Andreev intervalley scattering makes robust interference in clean 
\footnote{In disordered systems, other mechanisms contribute (see Refs.\ \cite{Manesco-QHgrS_SP22, KurilovichDisorderAES_NCom23} and discussion below) and produce irregular interference patterns which are key for interpreting the experiments.} 
zigzag nanostructures possible. 

For smooth or rounded corners, we use the result that almost all graphene edges act like a zigzag edge \cite{AkhBeenBoundaryPRB08, vanOstaayReconstructPRB11, Manesco-QHgrS_SP22} to generalize from our sharp corner results.  The key property is the type of effective zigzag edge present at the beginning versus at the end of contact with the superconductor.  The result for rounded corners is as if the incoming and outgoing modes were for these effective edge types. 

As the transparency of the interface, $T$, is reduced \cite{Transparency-def}, the Andreev conversion probability in the QH regime is 
%much 
more robust than that at zero field.  Despite the abrupt change in lattice, our numerical results for $P_{he}(T)$ at $B\!=\!0$ follow the classic result of Blonder, Tinkham, and Klapwijk (BTK) \cite{BTK-PRB82} obtained by matching free-electron wavefunctions: Andreev reflection decreases rapidly as the transparency is reduced.  In contrast, in the QH regime, Andreev conversion remains substantial, especially for small chemical potential, until the transparency drops below 10\%.  Thinking semiclassically, we attribute the robustness of Andreev conversion in the QH regime to multiple encounters with the superconductor due to the $e$-$h$ skipping orbits along the interface. 

This paper is organized as follows. In Sec.\,\ref{sec:InterfaceModel}, we first define the tight-binding model for the graphene-super\-con\-duc\-tor nanostructure in the context of the BdG equation.  Then, we give the typical parameter values used in this work (\ref{subsec:Parameters}), and review the reorganization of the energy spectrum from the usual QH edge states into chiral Andreev interface modes (\ref{subsec:Review-And.Modes}).  
Secs.\ \ref{sec:SimpleCases} and \ref{sec:ZigzagScattering} present scattering wave states for an incoming electron.  First, a zigzag parallelogram in two limiting cases is studied (\ref{sec:SimpleCases}): the interface is either fully opaque or fully transparent.  Then, we present scattering states for several zigzag nanostructures in which the transparency of the interface is reduced (\ref{sec:ZigzagScattering}).  
In Sec.\,\ref{sec:Conductance}, we turn to discussing Andreev conversion, or conductance, as a function of interface transparency, both at $B\!=\!0$ and in the QH regime.  Results for nanostructures with some non-zigzag edges are presented in Sec.\,\ref{sec:BeyondZigzagEdges},  including for those with rounded or smooth corners (\ref{subsec:SmoothCorners}).  We conclude in Sec.\,\ref{sec:Conclusion} and,
in particular, discuss the relevance of our results for experiments (\ref{subsec:Expts}).

\section{Model and Andreev Edge Modes}
\label{sec:InterfaceModel}

\subsection{Bogoliubov--de\,Gennes Approach}
\label{subsec:BdG-Approxes}

To study graphene-superconductor nanostructures, we use the same approach as in our previous study of the infinite, periodic graphene-superconductor interface \cite{AlexeyB_InfInter_PRB25}.  The standard mean field theory for spatially inhomogeneous superconductivity, the Bogoliubov--de Gennes (BdG) equation \citep{deGennesBook,BeenakkerRMP08,Asano_AndreevBook21}, yields the single-particle excitations of the system.  Introducing holes, one doubles the system's degrees of freedom and writes a single-particle real-space Hamiltonian in this doubled (Nambu) space.  Concretely, let $H(\mathbf{r})$ be the normal state single-particle Hamiltonian with one-body potential $U(\mathbf{r})$ and magnetic field described by the vector potential $A(\mathbf{r})$, and let $\Delta(\mathbf{r})$ be the superconducting pairing potential (the gap).  Then the BdG Hamiltonian is 
\begin{align}
\mathcal{H} & =\left(\begin{array}{cc}
H(\mathbf{r}) & \Delta(\mathbf{r})\\
\Delta^{*}(\mathbf{r}) & -H^{*}(\mathbf{r})
\end{array}\right),
\end{align}
where energy is measured from the chemical potential.  
%% The Nambu doubling builds in (artificial) particle-hole symmetry due to redundancy: if the wavefunction $(\Psi^{e}(\mathbf{r}),\Psi^{h}(\mathbf{r}))$ is a solution with energy $E$, then $(-\Psi^{h*}(\mathbf{r}),\Psi^{e*}(\mathbf{r}))$ is also a solution and has energy $-E$. 

The BdG equation is a convenient way to study the proximity effect between a superconductor and a normal system \cite{deGennesBook, BeenakkervanHouten-SuperQPC92,PannetierCourtoisJLTP00,Klapwijk_ProximityAndreev_2004,Asano_AndreevBook21, LesovikSadovskyyRev11}.  
The Cooper pairs described by the superconducting order parameter spill over causing electron pair correlations in the normal system.  These correlations are incorporated in the BdG approach, appearing as hybridization between electron and hole quasiparticles.  
Andreev reflection---the reflection of an incoming electron into an outgoing hole on the normal side 
\footnote{In Andreev reflection, one focuses on the excitations on the normal side of the interface.  These can be either an electron above the Fermi sea or the absence of an electron in the Fermi sea, namely a hole.  Individual excitations cannot propagate into the superconductor at the chemical potential; however, two electrons together can connect to the (nearly) zero momentum condensate in the superconductor. 
Thus, an incoming electron with wavevector $k$ at the interface pairs with a second electron with $-k$ and goes into the condensate.  The system therefore has a deficit of $-k$, which is described by an outgoing hole with wavevector $k$.  The resulting electron-hole hybridization describes the electron pairing correlations in the normal material induced by the superconductor (including a possible supercurrent).}---is 
the mechanism that establishes the proximity effect \cite{PannetierCourtoisJLTP00, Klapwijk_ProximityAndreev_2004}.  Andreev conversion is, then, a measure of the strength of the proximity effect in a nanostructure \cite{ProximityDecoherence}. 

In the mean-field BdG equation there are three fields that should be found self-consistently: the superconducting gap $\Delta(\mathbf{r})$, the one-body potential $U(\mathbf{r})$, and the vector potential $A(\mathbf{r})$.  We follow standard practice  
\citep{BTK-PRB82, TakagakiQHEPRB98,HoppeZulickePRL00, KhaymovichEPL10, Manesco-QHgrS_SP22, DavidGrenoble_Geometrical_PRB23, CuozzoRossiPRB24, LesovikSadovskyyRev11} in assuming that these fields are constant in each material and change abruptly at the graphene-superconductor interface. 
%, as discussed in our previous work on the infinite, periodic graphene-superconductor interface \cite{AlexeyB_InfInter_PRB25}. 
%We used this approach previously to study the infinite, periodic graphene-superconductor interface \cite{AlexeyB_InfInter_PRB25}.   
%We discuss this approach briefly here; further discussion can be found in our previous work on the infinite, periodic graphene-superconductor interface \cite{AlexeyB_InfInter_PRB25} in which we used the same approach.

The one-body potential $U(\mathbf{r})$ accounts for the abrupt change in Fermi energy at the interface [see the last term in Eqs.\,(\ref{eq:Ham-tbS-1})-(\ref{eq:Ham-tbQH-1}) below].  
%Since $\Delta(\mathbf{r})$ is certainly very small in graphene, we take it to be zero.  
$\Delta(\mathbf{r})$ can be taken to be zero in graphene (no attractive interaction), and
noting that the 2D graphene has little influence on pairing in the superconductor, we take $|\Delta|$ to be constant throughout the superconductor
\footnote{A step function in $|\Delta|$ at the interface is a good approximation for a 2D system meeting a 3D metallic superconductor \cite[{Sec.\,IV.A.2}]{Beenakker-RMT-RMP97} \cite{BeenakkerSpecularPRL06}.  First, the effective electron-electron attraction is likely to be so small in the 2D system that the pair interaction matrix element can be set to zero, $\Delta=0$.  Second, while $|\Delta|$ is usually suppressed in the superconductor near a normal system, here the density of states of the 2D system is much smaller than that of the superconductor, implying that the suppression is very small and can safely be neglected.  The suitability of this step-function form is supported by the classic analysis involving the conductivity of the normal and superconducting materials \cite[{Sec.\,IV.B.1}]{LikharevRMP79}.}. 
Although $\Delta(\mathbf{r})$ switches off abruptly at the interface, the superconducting order parameter does not switch off abruptly, as pairing correlations described by electron-hole hybridization extend into the graphene.  
Finally, as the superconductor naturally acts to screen the penetration of the magnetic field, we assume that $B$ drops abruptly from a constant value in graphene to zero in the superconductor.
%\footnote{In effect, we assume that $B$ is smaller than the lower critical field (i.e.\ the superconductor is essentially type I) and the London penetration depth is small}. 
While not the case in the recent experiments  \citep{LingfeiGlebCAESNatPhys20,LingfeiGlebLossPRL23,SahuDas_Noise_PRB21,HatefipourShabani_QH-S_NL22}, 
this approximation has nevertheless been used extensively \citep{TakagakiQHEPRB98, HoppeZulickePRL00, KhaymovichEPL10, ZhangTrauzettelPRL19, Manesco-QHgrS_SP22, DavidGrenoble_Geometrical_PRB23, CuozzoRossiPRB24, Panu-HeatChargePRB24}
to gain general information about the QH-superconductor interface because it is simple and enhances the effect of superconductivity. 
%Other work has made a very different simplifying approximation, namely to take $B$ constant across the interface with no decay  \citep{KurilovichDisorderAES_NCom23,MichelsenSchmidt_SupercurEnable_PRR23}.
%  which corresponds to a very large London penetration depth.

These approximations greatly simplify the problem: rather than solving the mean field theory self-consistently, one simply solves the BdG equation once for the fixed, constant values of 
% the potentials: 
$\Delta$, $B$, and the two Fermi energies. 
%The result is expected to be an overly optimistic picture of the influence of the superconductor on the QH state in graphene.  
Further discussion of these approximations can be found in our previous work on the infinite, periodic graphene-superconductor interface \cite{AlexeyB_InfInter_PRB25}. 

\subsection{Tight-Binding Model}
\label{subsec:TightBind} 

We consider a nanostructure in two dimensions described by the normal-state Hamiltonian   
\begin{equation}
H(x,y)=H_{S}+H_{QH}+H_{T},\label{eq:Ham-full}
\end{equation}
where $H_{S}$ represents the superconductor, $H_{QH}$ the QH graphene, and $H_{T}$ the connection between them. In the absence of spin-orbit coupling and the Zeeman effect, the problem decouples into two identical problems, leading to doubly degenerate states. We therefore consider just one spin sector and drop the spin index. 

\begin{figure}
\includegraphics[width=3.2in]{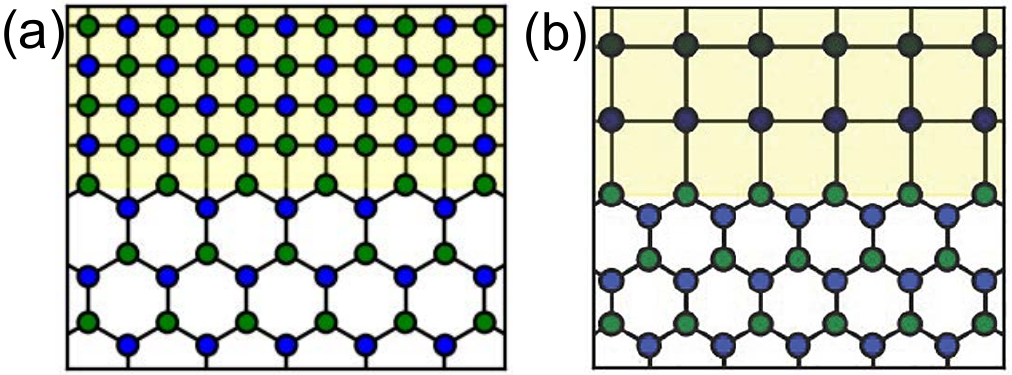}
\caption{Graphene-superconductor interface models: zigzag edge of graphene stitched to (a)~dense or (b)~sparse square lattice. The graphene sublattices---green (A) and blue (B)---extend into the superconductor for dense stitching. 
Atoms on the edge of the graphene sheet are in the A sublattice.} 
\label{fig:Stitching}
\end{figure}

The normal-state Hamiltonian of the superconductor is a square lattice in the $y>0$ half-plane (see Fig.\,\ref{fig:Stitching}),  
\begin{align}
H_{S}(x,y\!>\!0)= -t\underset{\langle ij\rangle}{\sum}\left(|i\rangle\langle j|+\textrm{h.c.}\right)
- (\mu_{S}\!-\!4t)\underset{i}{\sum}|i\rangle\langle i|,\label{eq:Ham-tbS-1}
\end{align}
where $\langle ij\rangle$ denotes nearest-neighbor sites. $\mu_{S}$ is the chemical potential (Fermi energy) measured from the bottom of the band. It appears as a constant because we neglect disorder (no onsite random potential) and electrostatic screening should be very good in the superconductor (no charge spill-over effects).  The pairing gap in the superconductor connects the electron states to holes on each site, 
\begin{align}
\Delta(x,y>0) = & \; \Delta\underset{i}{\sum}\left(|i\rangle_{e}\langle i|_{h}+\textrm{h.c.}\right).\label{eq:Delta-tb-1}
\end{align}

Graphene is modeled by hopping on a honeycomb lattice with phase factors due to the magnetic field: 
\begin{align}
H_{QH}(x,y\leq0)= & -t_{N}\underset{\langle ij\rangle}{\sum}\left(e^{-2\pi i\frac{\Phi_{ij}}{\Phi_{0}}}|i\rangle\langle j|+\textrm{h.c.}\right)\nonumber \\
 & -\mu_{\textrm{gr}}\underset{i}{\sum}|i\rangle\langle i|,\label{eq:Ham-tbQH-1}
\end{align}
where $\Phi_{ij}=\int_{r_{j}}^{r_{i}}\vec{A}\cdot d\vec{l}$ 
and $\Phi_{0}=h/e$ is the flux quantum. 
In the Landau gauge, $\vec{A}(x,y)=By\hat{x}\,\theta(-y)$ where $\theta(y)$ is the step function. With this choice, $\Delta$ is real and constant (in equilibrium) as in Eq.\,(\ref{eq:Delta-tb-1}). 
$\mu_{\textrm{gr}}$ is the chemical potential (Fermi energy) measured from the Dirac point.  Disorder is again neglected, and we assume as a simple model that electrostatic spill-over can be neglected as well. 

The graphene and superconductor edge sites are connected by hopping matrix elements of magnitude $t_{NS}$.  Since a zigzag edge is thought to be a good model for a
generic graphene edge \cite{AkhBeenBoundaryPRB08, vanOstaayReconstructPRB11, Manesco-QHgrS_SP22}, we assume for most of our work that the graphene ends in a zigzag edge.  (For non-zigzag results, e.g.\ armchair edges and rounded corners, see Sec.\,\ref{sec:BeyondZigzagEdges}.) We consider two ways of ``stitching'' the lattices together, shown in Fig.\,\ref{fig:Stitching}. 
First, in the ``dense'' stitching scenario {[}panel (a){]}, the square lattice of the superconductor has half the period of the graphene lattice and is connected to both graphene sublattices.  
Second, for ``sparse'' stitching {[}panel (b){]}, the square lattice has the same period as the zigzag edge and only the terminal sites are connected to the square lattice \citep{BlanterMartin_M-GrContacts_PRB07,TakagakiGrRibbons_JPCM21}.  

Our convention is that the terminal atoms of graphene at the interface belong to the A sublattice. To specify the hopping Hamiltonian connecting the graphene and superconductor lattices, we use graphene's lattice constant, $a$, as the unit of length in writing the $(x,y)$ coordinates of the edge sites. 
For sparse stitching [Fig.\,\ref{fig:Stitching}(b)], only the terminal A sites are connected to the square lattice, 
\begin{equation}
H_{T,\text{sparse}}= H_{T}^{(A)} \equiv -t_{NS}\underset{n}{\sum}^{(A)}\left[|n,\tfrac{1}{2}\rangle\langle n,0|+\textrm{h.c.}\right].
\end{equation}  
In the dense stitching scenario [Fig.\,\ref{fig:Stitching}(a)], the neighboring B sites are connected directly to the square lattice as well, 
\begin{align}
H_{T,\text{dense}}= & \,H_{T}^{(A)} \\
- &  \,t_{NS}\underset{n}{\sum}^{(B)}\left[|n+\tfrac{1}{2},\tfrac{1}{2}\rangle\langle n+\tfrac{1}{2},-\tfrac{\sqrt{3}}{2}|+\textrm{h.c.}\right].\nonumber 
\end{align}
Note that the density of states in the superconductor is larger for the dense scenario. 
%For instance, since the period of the square lattice is half that of the zigzag edge, the band structure of the square lattice should be folded back into the smaller graphene (projected) Brillouin zone.  

For this tight-binding Hamiltonian, we find the scattering states and transmission amplitudes through the nano\-structures using the Kwant quantum transport package \cite{GrothKwantNJP14}.  Since there is no particular symmetry relation between the honeycomb and square lattices that we use, nor between their Fermi surfaces (the superconductor is near half filling), we expect that general conclusions below will be valid for graphene-superconductor nano\-structures generically.  

This tight-binding model neglects several features of the experiments \citep{LingfeiGlebLossPRL23, LeeHarvardNatPhy17, LingfeiGlebCAESNatPhys20, SahuDas_Noise_PRB21, HatefipourShabani_QH-S_NL22} that motivate our study.  
First, we study a clean system as a basis for future work on disordered nanostructures.  While this is likely to be a good approximation for the graphene, the superconductor is typically an alloy and so heavily disordered.  
Second, we have neglected the Zeeman effect and spin-orbit coupling.  The effect of spin-splitting by itself should be small since it amounts to a small shift in energy and displacement of the interface wavefunction, as we found for an infinite interface \cite{AlexeyB_InfInter_PRB25}.  Spin-orbit coupling in the superconductor, however, could mix the two channels \cite{LeeHarvardNatPhy17, ZhangTrauzettelPRL19, MichelsenSchmidt_SupercurEnable_PRR23, Panu-HeatChargePRB24} and lead to new phenomena, which are outside the scope of this paper.  
Third, we do not include any effects of loss, decoherence, or temperature, which would make the proximity effect more complex and require extensions beyond the Andreev reflection considered here \cite{PannetierCourtoisJLTP00, Klapwijk_ProximityAndreev_2004, TangAlicea_VortexEnabled_PRB22,  SchillerOreg_FermDissip_PRB23, HuLian-Decohere-X24, LingfeiGlebLossPRL23}.  These are also left for future work.  
Further discussion of experimental relevance is given at the end of the paper (Sec.\,\ref{subsec:Expts}), after presentation of the results.

\subsection{Regime Studied and Parameter Values \label{subsec:Parameters}}

We work in a regime characterized by a hierarchy of energy scales, motivated by experiments \citep{LingfeiGlebCAESNatPhys20,LeeHarvardNatPhy17,LingfeiGlebLossPRL23,SahuDas_Noise_PRB21,HatefipourShabani_QH-S_NL22}.  The smallest scale is the superconducting gap, $\Delta$, and the largest scale is the bandwidth or Fermi energy in the superconductor. The magnetic energy is intermediate: we characterize it by the energy difference between the $n=0$ and $n=1$ Landau levels, denoted $E_L\equiv\sqrt{2}\hbar v/l_{B}$. In addition, the density of conduction electrons in graphene is much lower than in the superconductor. Thus we consider the regime  
\begin{equation} 
\Delta\ll E_{L},\mu_{\textrm{gr}}\ll \mu_{S}. 
\end{equation} 
In terms of length scales, the magnetic length ($l_{B}^{2}=\hbar/eB$) should be much larger than the lattice constant in either material. 

We use a standard set of parameter values for most of the results presented and note any deviations from these standard values. Our standard parameter values are $t_{NS}=t$, $E_{L}/t=0.12$, $\mu_{\textrm{gr}}/E_{L}=1/4$, and $\Delta/E_{L}=1/10$ 
\footnote{The width of the zigzag graphene nanoribbon is $\frac{300}{\sqrt{3}}a$. The width of the 
superconductor is, respectively, $150a$ and $400a$ for our dense and sparse stitching results.}. 
The magnetic length and magnetic field corresponding to this $E_{L}$ are $l_{B}/a=10.2$ and $Ba^{2}/\Phi_{0}=0.0015$ (or a flux of $0.0013\Phi_{0}$ per unit cell). In the superconductor, $\mu_{S}$ is close to half filling: our standard values are $\mu_{S}/t =4.5$ and $5.0$ for dense and sparse stitching, respectively. Thus the Fermi energy in graphene is much smaller than that in the superconductor: 
$\mu_{\textrm{gr}}/\mu_{S}\sim0.01$.

We obtain a well-matched, high-transparency \citep{Transparency-def} interface between the two lattices, and thus very similar scat\-ter\-ing-wave states, for a range of $\mu_{S}$ that is larger than the magnetic or superconducting energy scales, $E_{L}$ and $\Delta$, especially in the dense-stitched case. In our previous work \cite{AlexeyB_InfInter_PRB25}, we argue that the range is of order $t$.
%, the hopping matrix element. The oscillations and structure in the wavefunction in the superconductor are also discussed there. 
Unless otherwise noted, results in the main text are for dense stitching; analogous results for sparse stitching are given in the Supplemental Material \cite{SuppMat}.

\subsection{Andreev Interface Modes} %: Review 
\label{subsec:Review-And.Modes}

We start by briefly reviewing how QH edge states in a zigzag graphene nanoribbon reconfigure into chiral Andreev interface modes.  A more complete discussion of the 1D modes along an infinite periodic interface can be found in our previous work \cite{AlexeyB_InfInter_PRB25}. 

\begin{figure}
\includegraphics[width=3.2in]{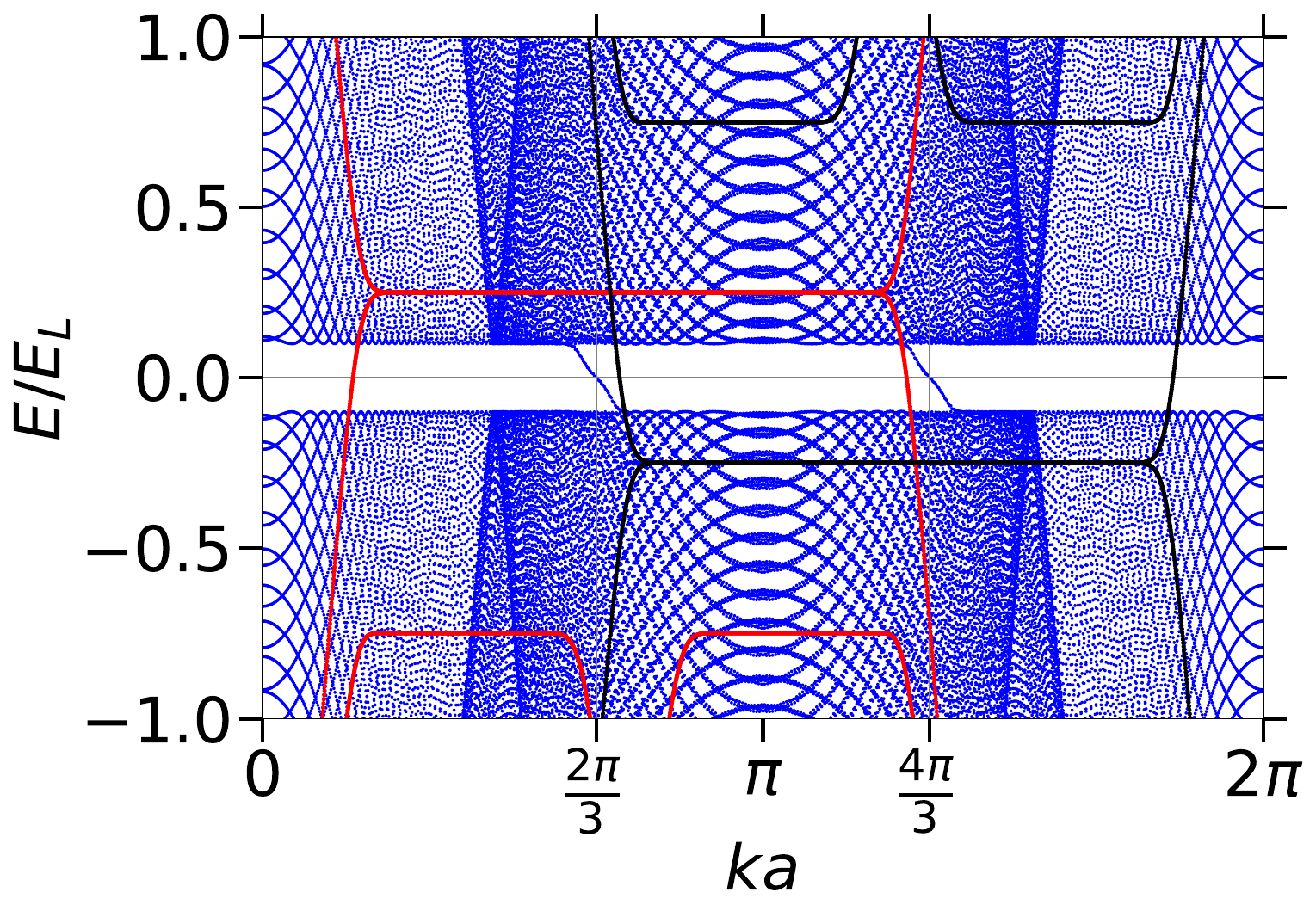}
\caption{The energy spectrum for graphene QH states [electron (black) and hole (red), $t_{NS}\!=\!0$] superposed with those for a well-matched, transparent graphene-superconductor interface (blue, $t_{NS}/t\!=\!1$ yielding $T\!\approx\!86\%$).  
The chiral Andreev interface modes  have energies within the superconducting gap and have $E\!=\!0$ at $\!k=\!K$ and $K'$ (marked by vertical lines).
(Dense stitching with standard parameters given in Sec.\,\ref{subsec:Parameters}.)} 
\label{fig:Dispersion}
\end{figure}

The BdG spectrum of an isolated zigzag nanoribbon in the QH regime is shown in Fig.\,\ref{fig:Dispersion}---both electrons (black lines) and holes (red lines).  It is convenient to choose the Brillouin zone as $ka\!\in\![0,2\pi]$.  The LLL states ($n\!=\!0$) for electrons and holes are the horizontal lines at $-\mu_\textrm{gr}$ and $+\mu_\textrm{gr}$ ($=\!E_L/4$), respectively, while the $n\!=\!1$ states can be seen at $E_L\!-\!\mu_\textrm{gr}$ (black) and $\mu_\textrm{gr}\!-\!E_L$ (red).  
Because of our choice of gauge and coordinate system, the electron state with guiding center coordinate at the interface ($y\!=\!0$) corresponds to wavevector $k=K=2\pi/3a$.  Likewise the hole state with guiding center at $y\!=\!0$ has $k=K'=4\pi/3a$.  
At the chemical potential ($E\!=\!0$), there are four propagating chiral modes: left-going particle and hole modes near $y\!=\!0$ and right-going particle and hole modes on the far side of the nanoribbon ($y\!=\!-W$).  For clarity, the width of the nanoribbon is chosen small enough so that the edge states on the far side appear in the same interval (the spectrum does not wrap around the Brillouin zone).

The modes near the $y\!=\!0$ edge, which we call the ter\-minat\-ed-lattice edge states, are clearly identified with a valley.  
The electron and hole states move in the same direction [the slope of $E(k)$ is the same].  Note that the $E\!=\!0$ propagating electron and hole states along a given zigzag edge cannot be coupled by superconductivity: 
they are in opposite valleys so there is a momentum mismatch (two electrons from the same valley cannot form a zero-momentum Cooper pair) [see, however, discussion of how screening current helps with momentum mismatch in Ref.\,\cite{MichelsenSchmidt_SupercurEnable_PRR23}]. 

We now attach a nanoribbon of superconductor to the graphene at $y\!=\!0$, schematically shown in Fig.\,\ref{fig:Geometry}, using dense stitching [Fig.\,\ref{fig:Stitching}(a)]. 
The BdG spectrum of this hybrid system is shown in blue.  For $|E|>\Delta$, the spectrum is dense because of the large density of states of the superconductor, and the Landau levels are no longer visible. 
%(it becomes a continuum in the limit that the superconductor occupies a half plane).  
For $|E|<\Delta$, the edge states on the far edge of the graphene nanoribbon are unchanged.  However, the spectrum for $k$ near $K$ and $K'$ is drastically modified, indicating that the interface states near $y\!=\!0$ are very different from the terminated-lattice edge states.  

The spectrum here is for a well-matched, transparent case so the interface modes have approximately equal electron and hole components in graphene \cite{AlexeyB_InfInter_PRB25}, 
indicating that Andreev reflection is strong. 
The velocity of these interface modes is greatly reduced compared to the terminated-lattice edge states \cite{AlexeyB_InfInter_PRB25} because the particles make long excursions into the superconductor.  Note that the modes at the chemical potential are at $k=K$ and $K'$, corresponding to guiding center coordinates at the interface, $y\!=\!0$.  In this case, the spectrum satisfies $E(k-K)=E(k-K')$ (for $|E|<\Delta$), and the spectrum is said to be ``valley degenerate''.

\section{Scattering States for Fully Opaque or Transparent Interface \label{sec:SimpleCases}}

\begin{figure}
\includegraphics[width=3.2in]{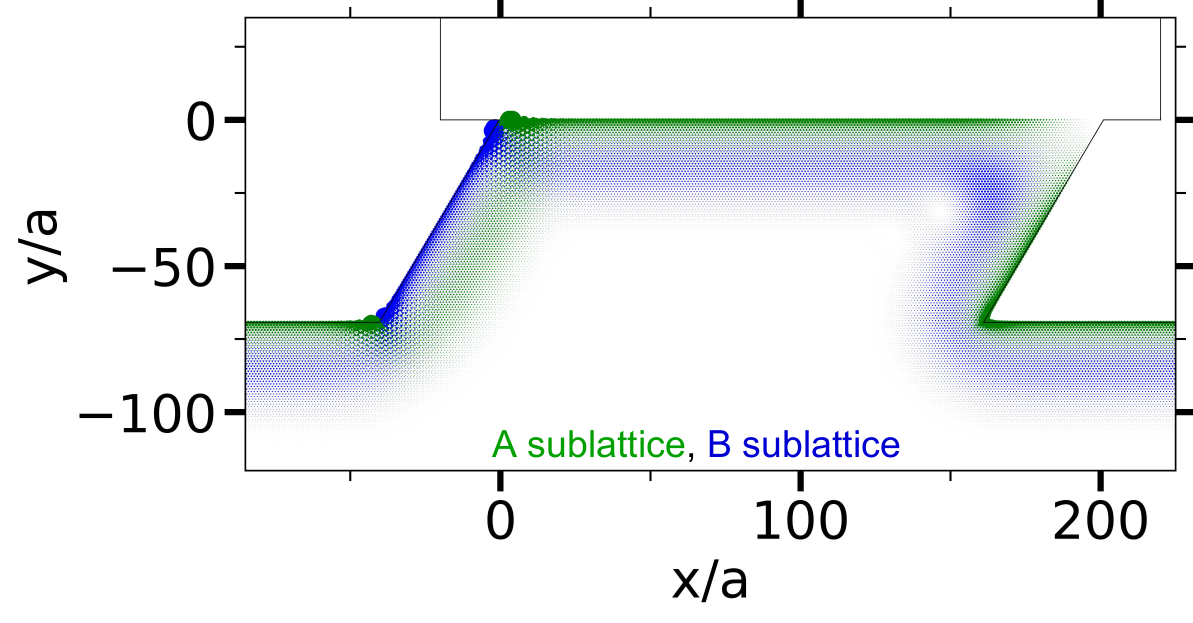}
\caption{Scattering state, $|\psi(x,y)|^2$, without superconductivity for graphene in the LLL in a 
zigzag parallelogram nano\-structure ($\theta_{1}\!=\!\theta_{2}\!=\!\pi/3$). 
Electrons are injected from the right in the $K$ valley for which the bulk wavefunction has amplitude only on 
the B sublattice (blue) and the terminal zig\-zag sites are on A (green). Note that the electron stays in the $K$ 
valley through the first two corners of the parallelogram but at the last two corners (left side) undergoes 
strong intervalley scattering.  ($t_{NS}\!=\!0$, otherwise standard parameters in Sec.\,\ref{subsec:Parameters}; the plotting method is explained in \cite{PlottingMethod}.)
}
\label{fig:DisconScatState}
\end{figure}

\begin{figure}
\includegraphics[width=3.2in]{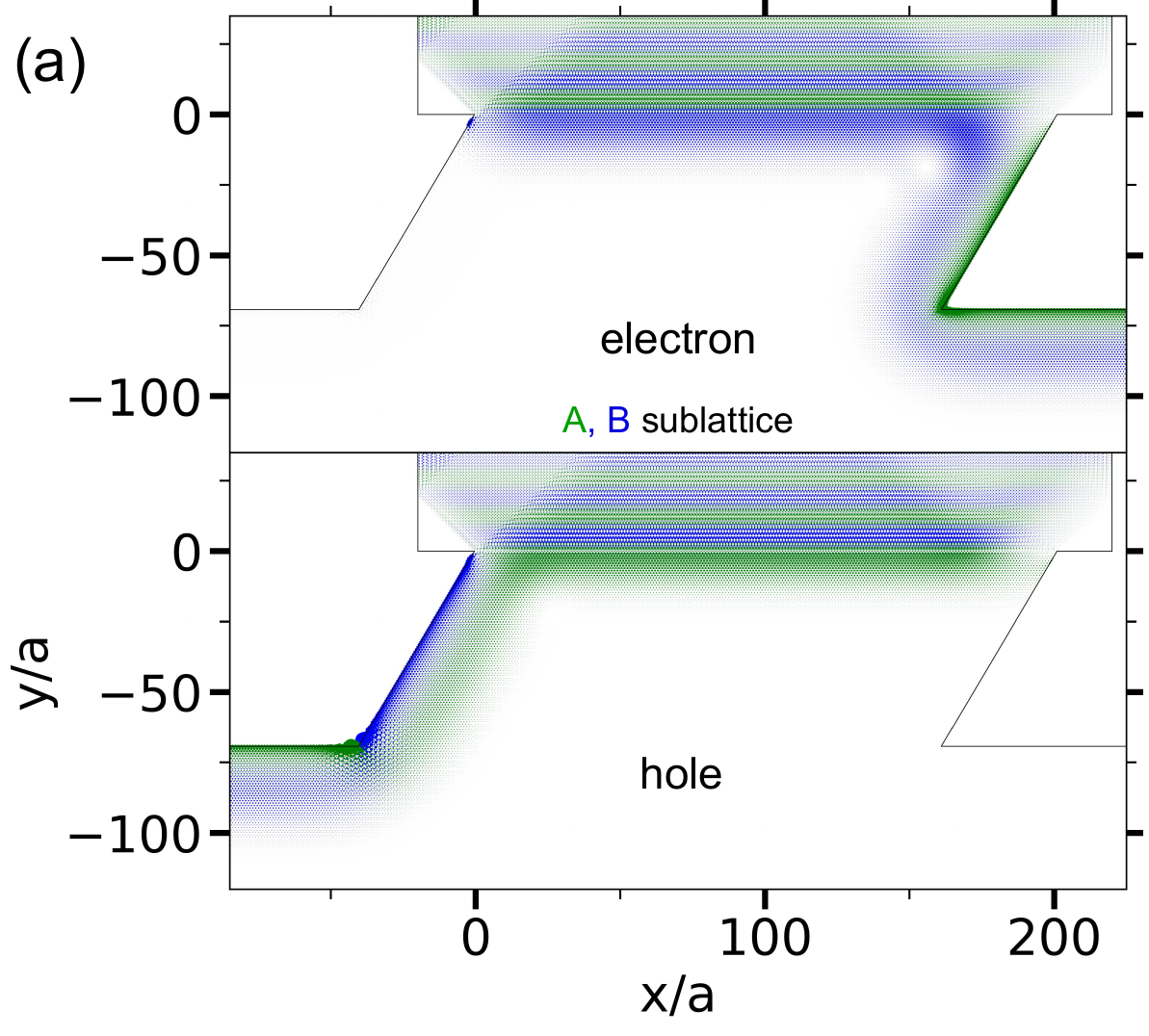}
\includegraphics[width=3.2in]{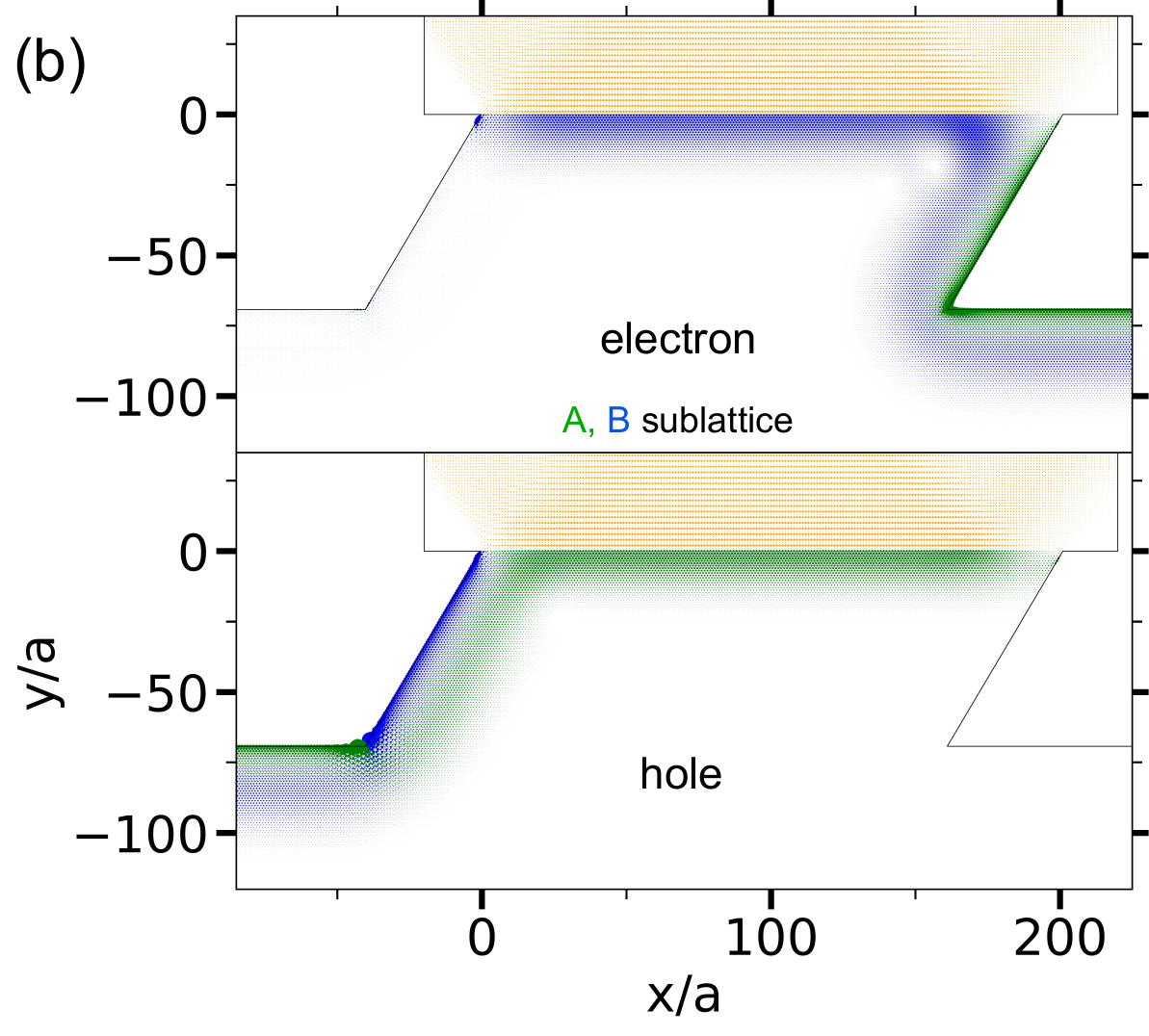}
\caption{Scattering state, $|\psi^{\beta}(x,y)|^2$ with $\beta=e,h$, for a zigzag parallelogram nanostructure at full graphene-superconductor coupling  and (a) dense or (b) sparse stitching. 
In each panel the electron (hole) component is in the upper (lower) part. 
%A parallelogram nanostructure is used with $t_{NS}=t$, $\mu_\textrm{gr}/E_L=1/4$ and, to ensure a transparent interface, $\mu_{S}/t=3.8$ and $5.0$, respectively \cite{AlexeyB_InfInter_PRB25}. 
Note the smooth connection from the interface state to the hole state along the left-hand side, leading to nearly perfect 
Andreev conversion in both cases. 
(Standard parameters in Sec.\,\ref{subsec:Parameters}; the plotting method is explained in \cite{PlottingMethod}.)}  
\label{fig:ScatterState_parallelogram} 
\end{figure}

The properties of the infinite-interface hybrid modes strongly influence the scattering states in nanostructures.  
We start by showing scattering states for two simple examples. The structure is a parallelogram in which all the graphene edges are zigzag [$\theta_{1}=\theta_{2}=\pi/3$ in the geometry of Fig.\,\ref{fig:Geometry} (red lines)].  
The graphene and superconductor are either completely disconnected (Fig.\,\ref{fig:DisconScatState}) 
or connected as well as possible (Fig.\,\ref{fig:ScatterState_parallelogram}). 

In both figures, the incoming state enters the scattering region from the right and is the electron QH edge mode near $k=K$. The outgoing state at the left side of the structure can be either electron or hole (or a combination). 

The plotting method used here and throughout the paper is as follows. At each lattice site $i$, we plot a disk with area proportional to $|\psi^\beta_i|^2$ ($\beta=e,h$). The disk is green for sublattice A sites and blue for sublattice B.  The intensity scale is chosen such that 
the disks on the terminal A sites of the incoming electron state just barely overlap  (for our standard parameters given in Sec.\,\ref{subsec:Parameters}).  
%Disks are smaller and non-overlapping at higher doping $\mu_\textrm{gr}/E_L=1/2$ since the surface state is less localized to the boundary \citep{BreyFertig_EdgePRB06}.  
Note that there is saturation in the wavefunction peaks at two of the corners.

Sublattice-valley locking is a key feature of the LLL states in graphene. Through this connection, the local sublattice structure reveals the valley content of the state (that is, whether $k$ is near $K$ or $K'$) at that position in the nanostructure. Our convention is that a bulk $K$-valley state has weight on only the B sublattice.
%, colored blue in the figures. 
In the incoming edge mode in Fig.\,\ref{fig:DisconScatState}, the terminated lattice boundary condition causes there to be equal weight on the two sublattices---that of the termination atoms (A, green) and that of the bulk LLL state (B, blue). (For more 
about QH edge states see \cite{BreyFertig_EdgePRB06,DelplaceMontambauxPRB10,Romanovsky_DiracQH_PRB11}.)

In the absence of the superconductor, Fig.\,\ref{fig:DisconScatState}, the scattering state is, of course, entirely electron. From the sublattice structure, one can follow the scattering of the electron between the $K$ and $K'$ valleys as it moves through the nanostructure. As the particle comes in from the right, at the first two corners the sublattice structure stays the same. Thus the particle turns these corners while remaining in the same valley. 
In contrast, at the third and fourth corners (left side of parallelogram), the weight on the A and B sublattices are exchanged. This indicates a change of valley, implying a large momentum scattering event at the corner caused by the abrupt change on the lattice scale. This intervalley scattering is associated with spikes of the wavefunction at the third and fourth corners. Indeed, because the QH edge state on the left side is in the $K'$ valley (terminal atoms are blue indicating B sublattice) 
\footnote{The left and right sides of the parallelogram in Fig.\,\ref{fig:DisconScatState} form a graphene nanoribbon. For opposite sides of a zigzag nanoribbon, the terminal sites are on opposite sublattices and so the LLL edge states are in opposite valleys.}, 
intervalley scattering at the third and fourth corners is required.  

The striking difference in the effect of a $60^\circ$ corner (left-hand corners in Fig.\,\ref{fig:DisconScatState}) compared to a $120^\circ$ corner (right-hand corners), where the angle is the change in direction of propagation, follows from the point group symmetry of the honeycomb lattice. When a honeycomb lattice is rotated by $60^\circ$, the A and B sublattices are interchanged, and the sublattice of the terminal sites changes. 
Sublattice-valley locking then implies that the valley of the propagating edge state changes, forcing intervalley scattering. We call this an \emph{intervalley-scattering} corner. 
In contrast, $120^\circ$ rotations preserve the sublattice of the zigzag edge and hence the valley of the edge state: large momentum scattering is not needed. This is a \emph{valley-preserving} corner. 
% Note that the wavefunction at these corners (see Fig.\,\ref{fig:DisconScatState}) is remarkably smooth. 

Turning now to high transparency graphene-super\-con\-duc\-tor interfaces, Fig.\,\ref{fig:ScatterState_parallelogram}, 
we show typical scattering-wave states for both dense and sparse stitching 
($t_{NS}/t\!=\!1$ yielding $T\!\approx\!85\%$ \citep{Transparency-def}). 
In both cases, the output consists 
%almost 
entirely of holes: consistent with 
Eq.\,(\ref{eq:Phe-ideal-AkhBeen}), the Andreev conversion is \emph{perfect}. The sublattice structure of the 
outgoing  hole is the same as that of the incoming electron, as it must be since the propagating hole is simply the 
absence of a propagating electron. 
% As noted above (Fig.\,\ref{fig:Dispersion}), 
% the hole is in the opposite valley ($K'$) from the incoming electron ($K$). 

The wavefunction along the interface is clearly very different from the 
terminated-lattice wavefunction (compare Figs.\,\ref{fig:DisconScatState} and \ref{fig:ScatterState_parallelogram}). The particle amplitude in graphene is mostly 
on sublattice B (the sublattice for the bulk $K$-valley state) while the hole amplitude is on sublattice 
A, which corresponds to a bulk hole in the $K$-valley (absence of a $K'$ electron). Thus, this is the $K$-valley 
electron-hole hybrid state identified in the spectrum of Fig.\,\ref{fig:Dispersion}. (For discussion of the wavefunction in the superconductor, see Ref.\,\cite{AlexeyB_InfInter_PRB25}.) Indeed, the intensity is uniform 
along the interface (away from the corners): the absence of a beating pattern indicates the presence of a single 
electron-hole hybrid mode in the scattering state. 

The connections in to and out of the interface are remarkably smooth. At the first two corners of the parallelogram, the incoming electron 
smoothly turns the corners, remaining in the $K$ valley. Thus we see that the incoming 
electron edge state smoothly connects to either the $K$-valley QH edge state (Fig.\,\ref{fig:DisconScatState}) or the $K$-valley electron-hole  hybrid state along the interface (Fig.\,\ref{fig:ScatterState_parallelogram}), depending on the circumstances. 

At the third corner, the hole wavefunction along the interface ($K$ valley) smoothly connects to the propagating 
hole along the left side of the parallelogram, which is also in the $K$ valley. 
(The propagating electron state is in the $K'$ valley.)
Thus, no intervalley scattering need occur at the third corner, and the absence of any increased 
wavefunction amplitude confirms that intervalley scattering is weak. At the final corner of the parallelogram, the 
hole must scatter, as in Fig.\,\ref{fig:DisconScatState}, to the other valley ($K'$) in order to propagate along the output edge.

%\section{Scattering States in Zigzag Nanostructures: Partially Transparent Interface
\section{Partially Transparent Interface in Zigzag Nanostructures}
\label{sec:ZigzagScattering}

\subsection{Parallelogram: One Intervalley Corner at Entrance or Exit
\label{subsec:Parallelogram}}

Reducing the transparency \citep{Transparency-def} of the graphene-super\-con\-ductor interface must lead to a crossover from the scattering state in  Fig.\,\ref{fig:ScatterState_parallelogram} to that in Fig.\,\ref{fig:DisconScatState}.  We study this crossover from nearly perfect Andreev conversion to complete absence by varying the hopping matrix element between graphene and the superconductor, $t_{NS}$. The resulting scattering-state wavefunctions show the crucial role played by intervalley-scattering corners. 

\begin{figure}
\includegraphics[width=3.2in]{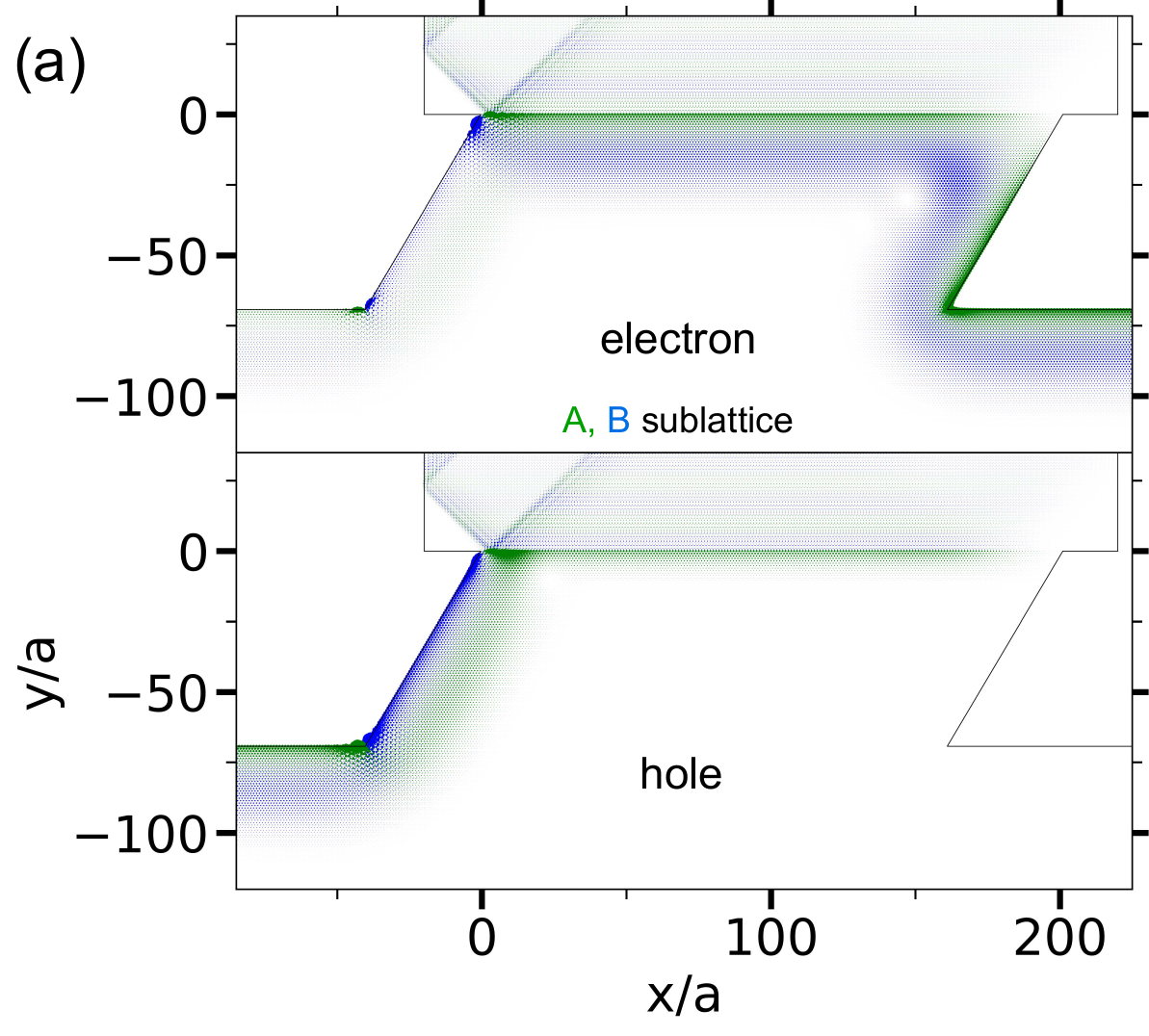}
\includegraphics[width=3.2in]{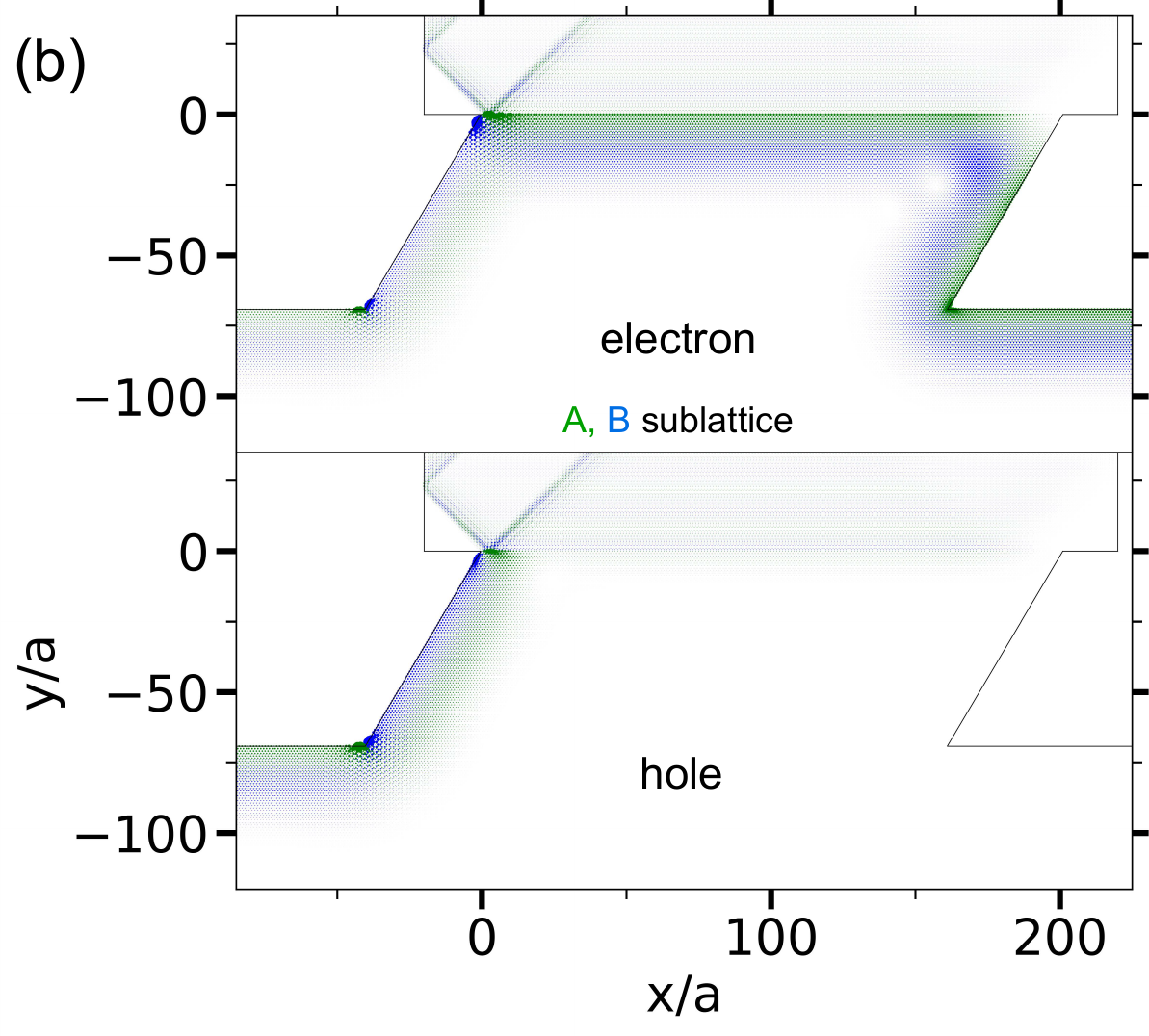}
\caption{Scattering state, $|\psi^{\beta}(x,y)|^2$ with $\beta\!=\!e,h$, for a zig\-zag parallelogram at reduced interface transparency ($\approx\!23\%$, $t_{NS}/t\!=\!0.3$) and 
(a) low filling, $\mu_\textrm{gr}/E_L\!=\!1/4$, or (b)~mid-Landau-gap, $\mu_\textrm{gr}/E_L\!=\!1/2$. 
In each panel the electron (hole) component is in the upper (lower) part \cite{PlottingMethod}. 
Despite weak electron-hole hybridization along the interface, the hole component dominates the output, produced by Andreev intervalley scattering at the downstream corner of the interface (left-hand side). 
(Parameters otherwise standard, see Sec.\,\ref{subsec:Parameters}.)  
\label{fig:Paralel1-reduced}
}
\end{figure}

In the typical zigzag parallelogram structures shown in Fig.\,\ref{fig:Paralel1-reduced}, the injected electron maintains its valley through the first two corners, as in the high-transparency and disconnected cases above.  
The electron smoothly connects to the hybrid mode along the interface that is in the same valley: the scattering state along the interface is smooth and has the periodicity of the lattice.  
Compared to the fully transparent interface (Fig.\,\ref{fig:ScatterState_parallelogram}), the degree of hybridization here is greatly reduced.

\begin{figure}
\includegraphics[width=3.2in]{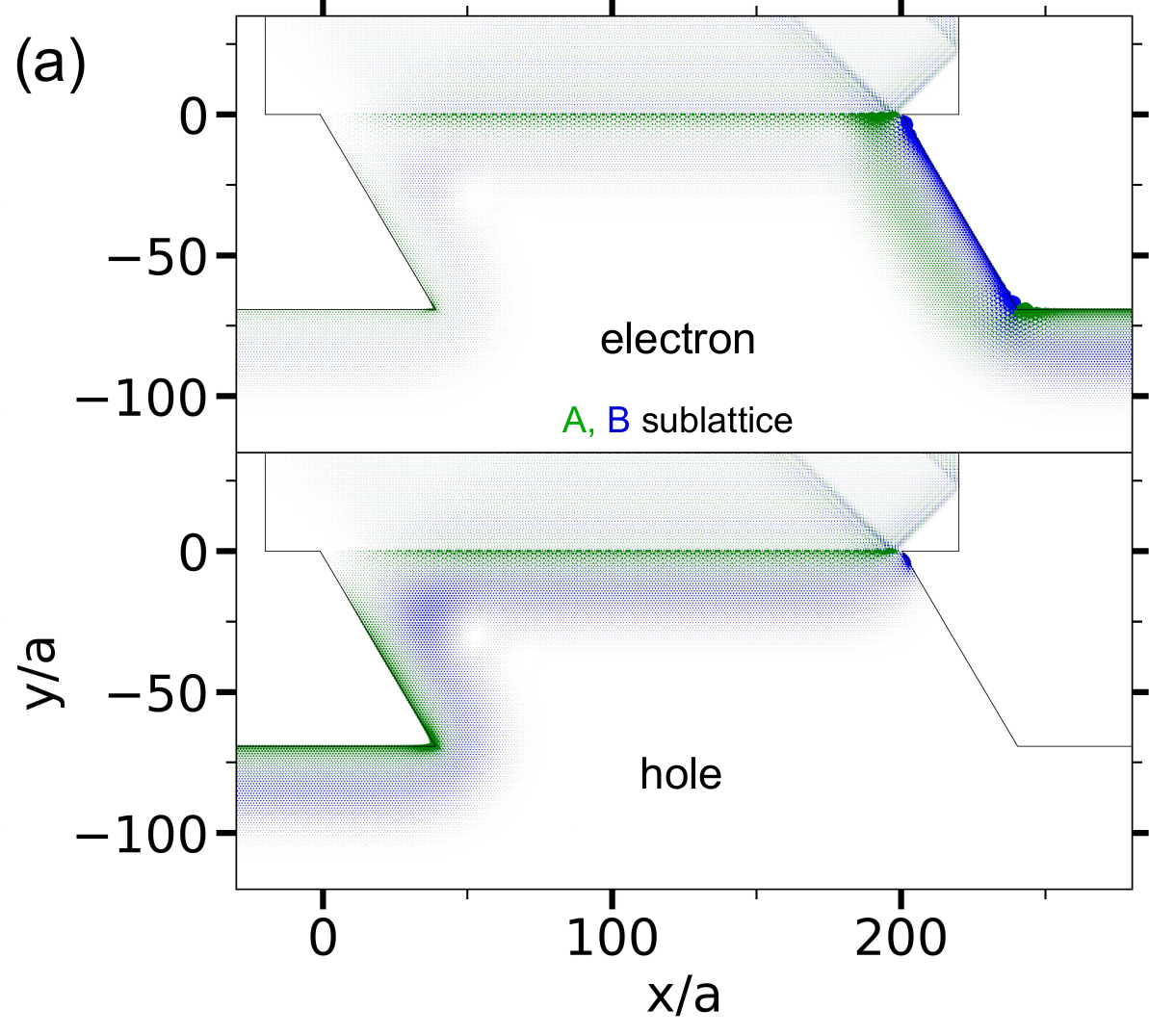}
\includegraphics[width=3.2in]{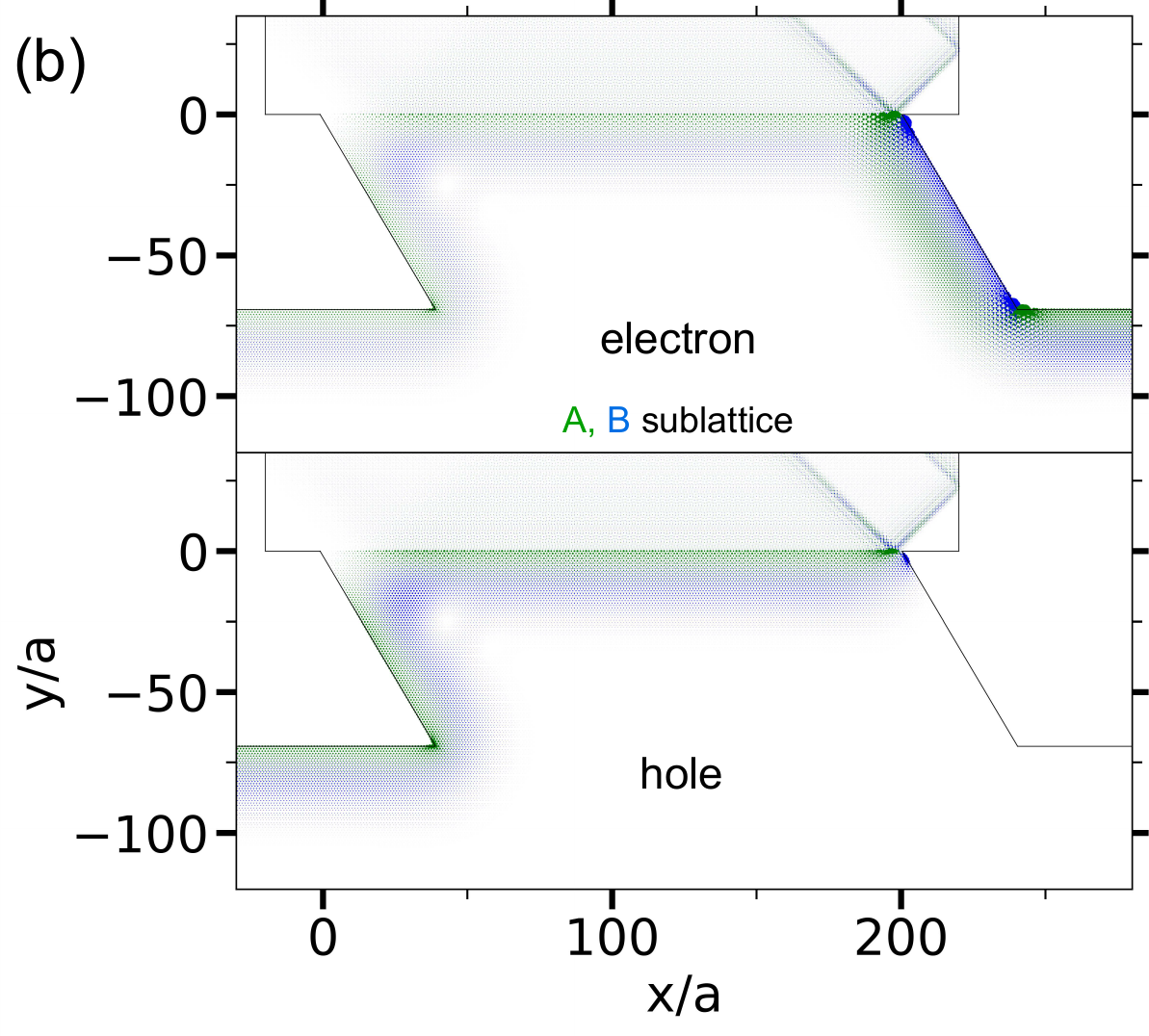}
\caption{Scattering state $|\psi^{\beta}(x,y)|^2$ at reduced interface transparency, as in Fig.\,\ref{fig:Paralel1-reduced} but with the orientation of the zigzag parallelogram reversed.  Because the intervalley-scattering corner is now upstream of the interface (right-hand side), both interface modes are populated (note beating).  
The final output intensity is independent of the parallelogram's orientation. 
%  The stronger conversion at the lower doping (a) remains because the scattering state consists of the modes with better interface hybridization than at higher doping (b).  
($t_{NS}/t\!=\!0.3$, (a) $\mu_\textrm{gr}/E_L\!=\!1/4$, (b) $\mu_\textrm{gr}/E_L\!=\!1/2$; parameters otherwise standard, Sec.\,\ref{subsec:Parameters}.)} 
\label{fig:Paralel2-reduced}
\end{figure}

At the end of the interface (third corner), the local increase in the wavefunction indicates that intervalley-scattering occurs.   
A novel feature here is that additional coupling to the superconductor is triggered by the intervalley-scattering corner.  This \emph{Andreev intervalley scattering} accompanies any local spike in the wavefunction since the evanescent waves present connect to energies above the superconducting gap where graphene and superconductor states are hybridized.  
Remarkably, a hole state is preferentially produced at the exit,  
which then propagates along the left side of the parallelogram.  
(Similar results for a sparse-stitched interface are shown in the supplemental material \cite{SuppMat}.) 
The net result is as if the hybridization of the interface mode were higher. 

When the parallelogram is reversed as in Fig.\,\ref{fig:Paralel2-reduced}, intervalley scattering occurs as the injected electron meets the interface. Again, enhanced coupling to the superconductor accompanies the intervalley scattering. However, both interface modes are excited: note the beating pattern in the wavefunction along the interface (seen more clearly in the trapezoid in the next subsection). Thus the comparable electron and hole weight along the interface here is \emph{not} an indication of good $e$-$h$ hybridization.  Rather, it indicates a comparable weight in the mostly $e$-like and mostly $h$-like modes, where the weight in the $h$-like mode comes from Andreev intervalley scattering.  The corner on the left side of the interface preserves the valley structure. Thus, the interface mode in the $K$ valley connects to an outgoing electron in the $K$ valley, while the interface mode in the $K'$ valley connects to an outgoing hole, which is necessarily in the $K'$ valley. 

The transmission and Andreev conversion probability must be the same in the two structures, because the chiral state in one orientation is related to that in the other by time reversal. 
Figs.\,\ref{fig:Paralel1-reduced} and \ref{fig:Paralel2-reduced} show, however, that the wavefunctions are quite different (even at full coupling 
\footnote{At full coupling, $t_{NS}/t\!=\!1$, nearly perfect Andreev conversion
occurs, which implies that only the interface mode in the $K'$ valley
is involved for the reverse parallelogram [electron on A sublattice (green) and hole on B (blue)].  In contrast, only the $K$-valley interface mode appears in Fig.\,\ref{fig:ScatterState_parallelogram}.}). 
Even though both interface modes are excited in the reverse parallelogram, no quantum interference occurs in the transmission since each mode is transmitted independently from the interface to the outgoing edge.  

\begin{figure}
\includegraphics[width=3.2in]{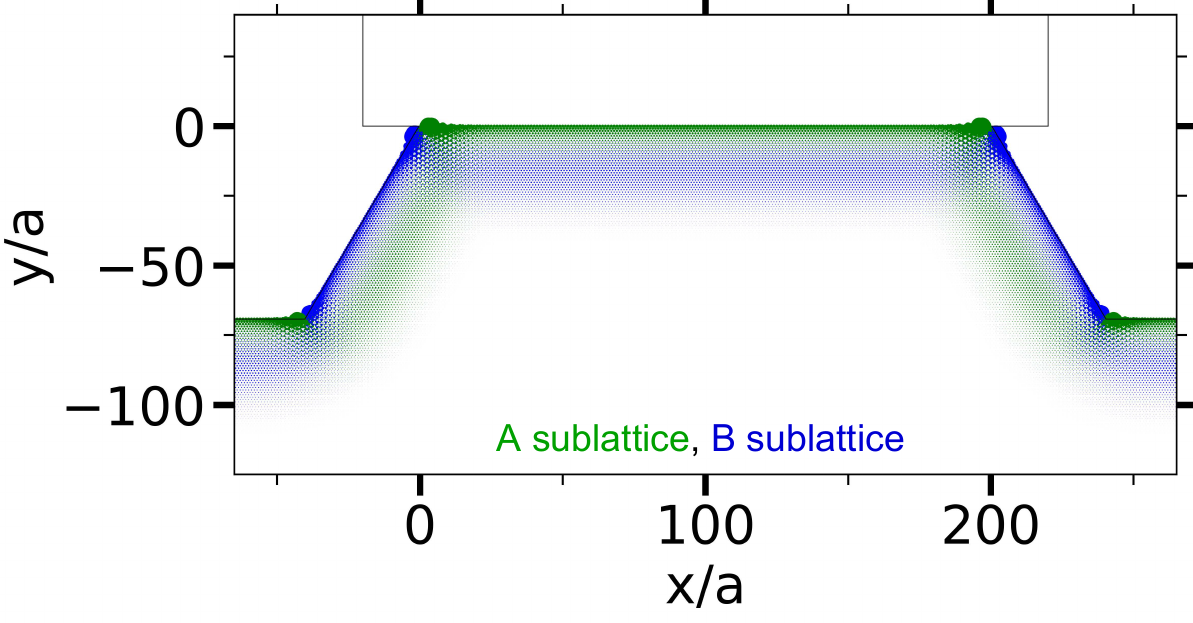}
\caption{Scattering state, $|\psi(x,y)|^2$, without superconductivity for a zigzag trapezoid nanostructure  
($\theta_{1}\!=\!2\pi/3$, $\theta_{2}\!=\!\pi/3$). Note that there is intervalley scattering at all four corners. 
($t_{NS}\!=\!0$, parameters otherwise standard, Sec.\,\ref{subsec:Parameters}.) 
} 
\label{fig:Trap-noS} 
\end{figure}

\begin{figure}
\includegraphics[width=3.2in]{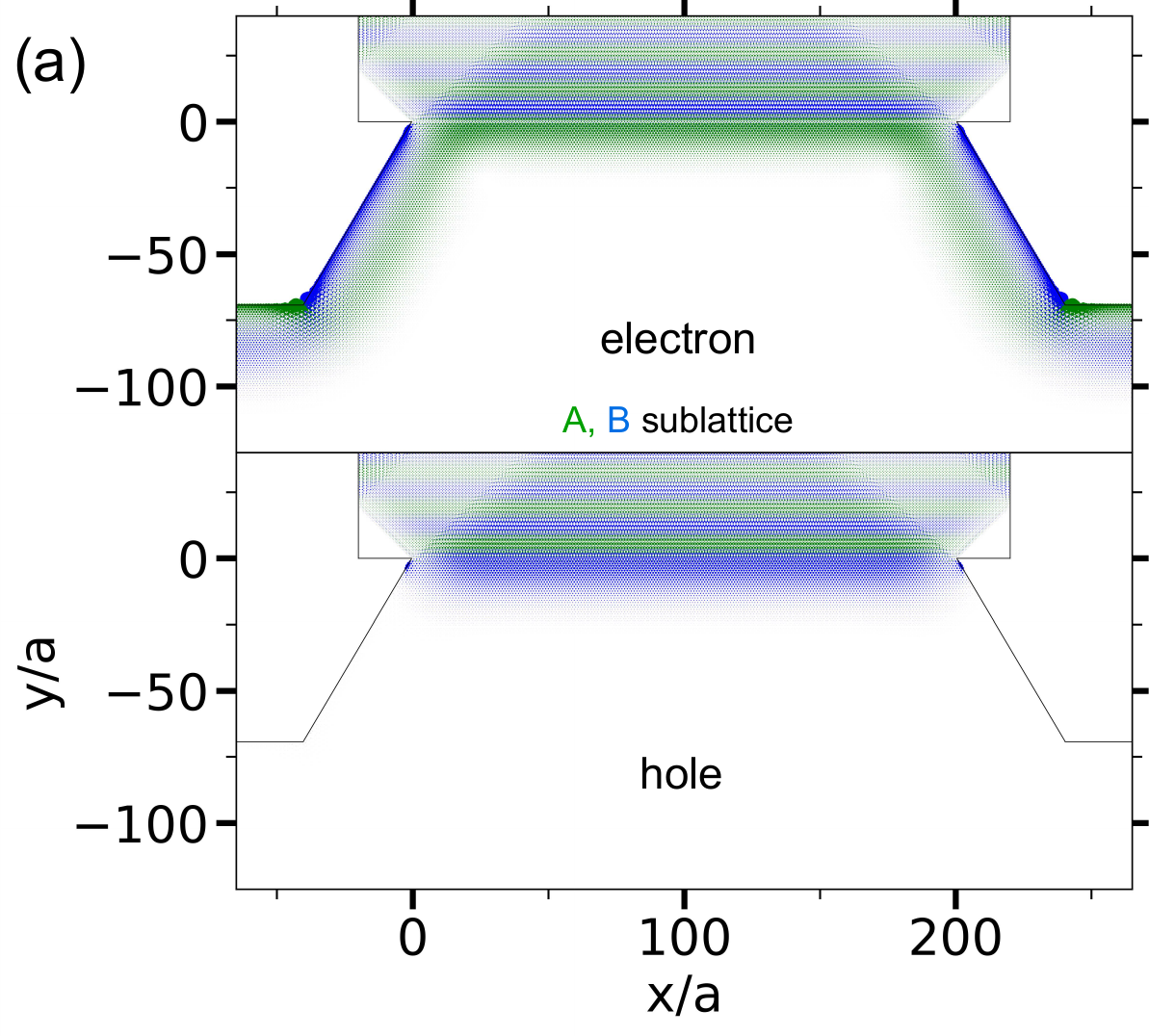}
\includegraphics[width=3.2in]{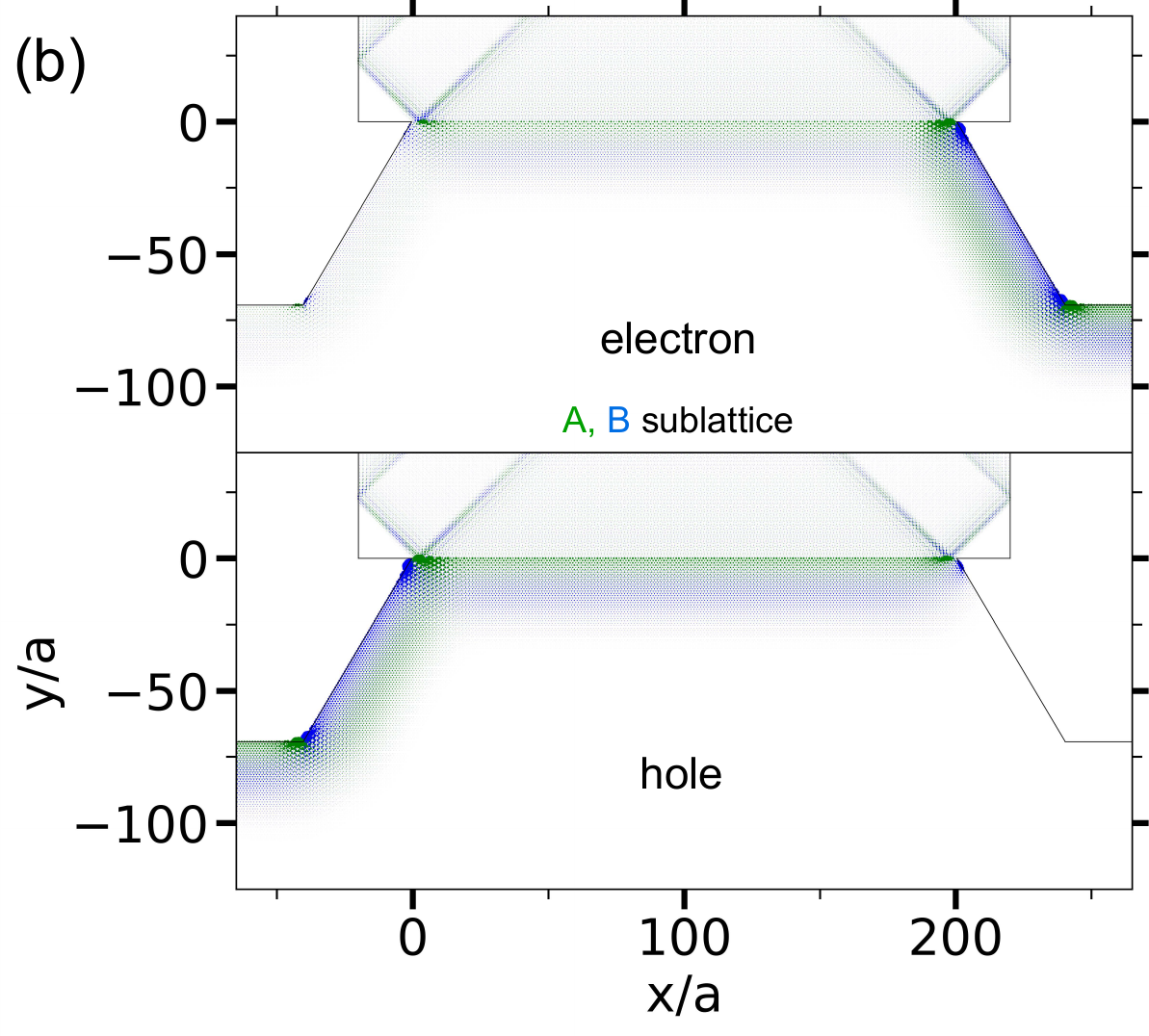}
\caption{Scattering state, $|\psi^{\beta}(x,y)|^2$ for $\beta\!=\!e,h$,  for a zigzag trapezoid nanostructure at (a)~full or (b)~partial transparency ($\approx\!23\%$ \citep{Transparency-def}).  In each panel the electron (hole) component is in the upper (lower) part \cite{PlottingMethod}. 
For full transparency, intervalley scattering is absent. Note that the hybrid interface mode is opposite to that for the parallelogram (compare Fig.\,\ref{fig:ScatterState_parallelogram}).  At partial transparency, intervalley scattering occurs at both interface corners, leading to interference effects. 
[$t_{NS}/t\!=\!0.3$ and $\mu_\textrm{gr}$ is for the peak near $\mu_\textrm{gr}/E_L\!=\!0.5$ in Fig.\,\ref{fig:SummaryPA}(b) (green line); parameters otherwise standard (Sec.\,\ref{subsec:Parameters}).] 
} 
\label{fig:Trap-withS} 
\end{figure}

\subsection{Trapezoid: Two Intervalley Corners \label{subsec:Trapezoid}}

In order to have intervalley-scattering corners at both the input and output of the 
interface, we next consider a trapezoid (green lines in Fig.\,\ref{fig:Geometry}). The scattering wavefunction in the absence of superconductivity is shown in Fig.\,\ref{fig:Trap-noS}. Since all four corners are $60^\circ$, intervalley scattering occurs at every corner. The change in valley of the propagating electron state at each corner is reflected in the change in sublattice structure: the weight on the A and B sublattice interchange at each corner.  

Two cases with superconductivity are shown in Fig.\,\ref{fig:Trap-withS}: when the interface has full or reduced transparency. At full transparency, since the valley of the propagating electron state is the same on the left and right edges of the trapezoid, there is a smooth connection to the hybrid interface state that is in the same valley (namely $K'$). 
There is, thus, no net Andreev conversion: $P_{he}\!=\!0$, consistent with Eq.\,(\ref{eq:Phe-ideal-AkhBeen}). Note that this interface state is opposite to the one in the full-transparency parallelogram, Fig.\,\ref{fig:ScatterState_parallelogram}. Indeed, the electron part of the interface state here corresponds to a non-propagating state for this zig\-zag edge, providing a clear illustration that large $t_{NS}$ reorganizes the states along the interface.  

\begin{figure}
\includegraphics[width=2.8in]{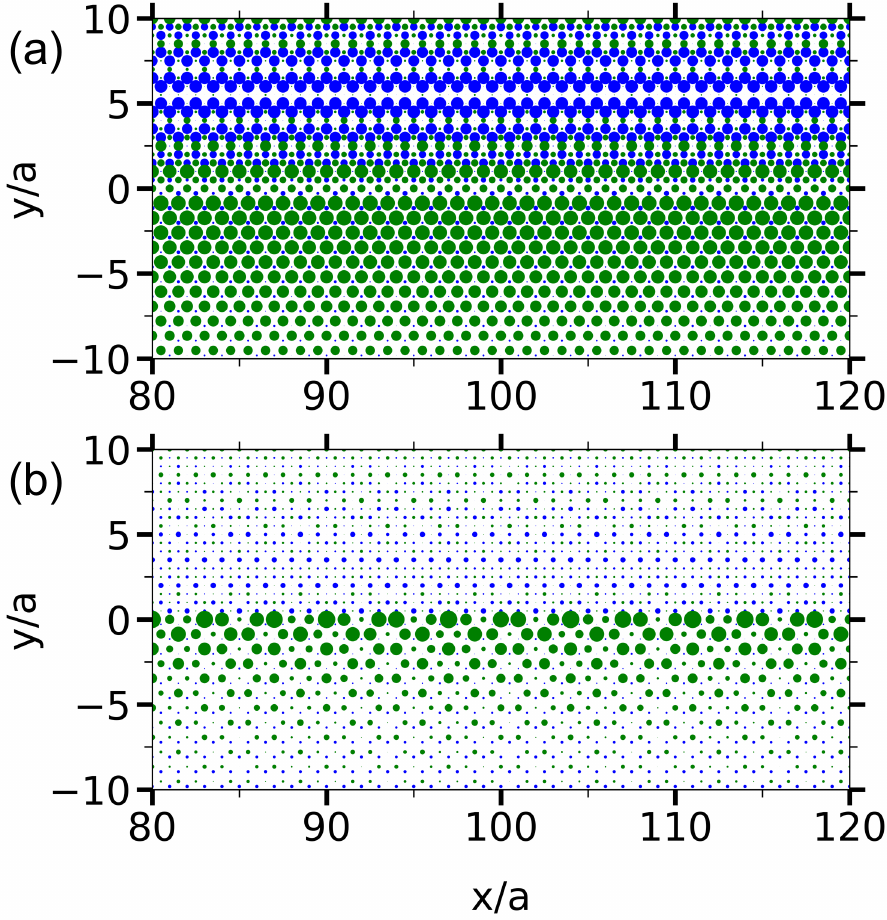}
\caption{ Detail of the electron component $|\psi^e(x,y)|^2$ at the two interfaces in Fig.\,\ref{fig:Trap-withS}: (a)~full transparency, (b)~partial transparency. The beating in panel (b) indicates that both interface modes are occupied, while the lack of beating in panel (a) shows that a single mode is present. The beating pattern is similar to that in Fig.\,\ref{fig:Paralel2-reduced} for the parallelogram.}
\label{fig:beating-trap} 
\end{figure}

In contrast, for reduced (but non-zero) transparency, Fig.\,\ref{fig:Trap-withS}(b), there is intervalley scattering at both ends of the interface. 
%(The interface transparency in this case is $\approx\!23\%$.) 
Note that both modes along the interface are excited: the zoom in Fig.\,\ref{fig:beating-trap}(b) shows the resultant beating. Interference between the two modes causes $P_{he}$ to become sensitive to parameters (e.g.\ $\mu_\textrm{gr}$, $B$, and $L$). Eq.\,(\ref{eq:Phe-ideal-AkhBeen}) no longer applies: for example, the periodic oscillation of $P_{he}$ as the graphene chemical potential is scanned across the LLL is shown for this case in Fig.\,\ref{fig:SummaryPA}(b).  

Andreev intervalley scattering is evident in Fig.\,\ref{fig:Trap-withS}(b) at both corners through the weight in the superconductor emanating from the corners. As a result, the outgoing particle is mostly in the hole sector, in contrast to the outgoing electron for \emph{both} zero and full transparency ($t_{NS}/t=0$ or $1$). 
Andreev intervalley scattering leads to strong oscillations in  $P_{he}$ despite the weak electron-hole hybridization along the interface. We argue later that this is not 
specific to the zigzag trapezoid 
but occurs without fine-tuning the angles defining the structure. 

\subsection{Inverted Trapezoid: Zero Intervalley Corners}
\label{subsec:InvertTrap}

The final all-zigzag-edge structure that might be examined is an inverted trapezoid, 
$\theta_1=\pi/3$ and $\theta_2= 2\pi/3$ in the geometry of Fig.\,\ref{fig:Geometry},  
in which the graphene gets narrower as one moves away from the interface (not shown). Here all four corners are $120^\circ$ and so there is no intervalley scattering necessary at any coupling to the superconductor. The output is always in the electron sector, $P_{he}=0$, for all values of $t_{NS}$.

\section{Conductance: Chiral Andreev Conversion}
\label{sec:Conductance}

Having illustrated the main effects with several examples, we now study more systematically how  Andreev conversion depends on both the interface transparency and the graphene chemical potential.  The probability of chiral Andreev conversion through the nanostructure is directly related to its conductance, 
%$G\!=\!(2e^2/h) \left[1-2P_{he}\right]$. 
\begin{equation}
%    G = (2e^2/h) \left[1-2P_{he}\right]
    G = \frac{2e^2}{h} \left(1-2P_{he}\right) \;.
\end{equation}

\begin{figure}
\includegraphics[width=3.0in]{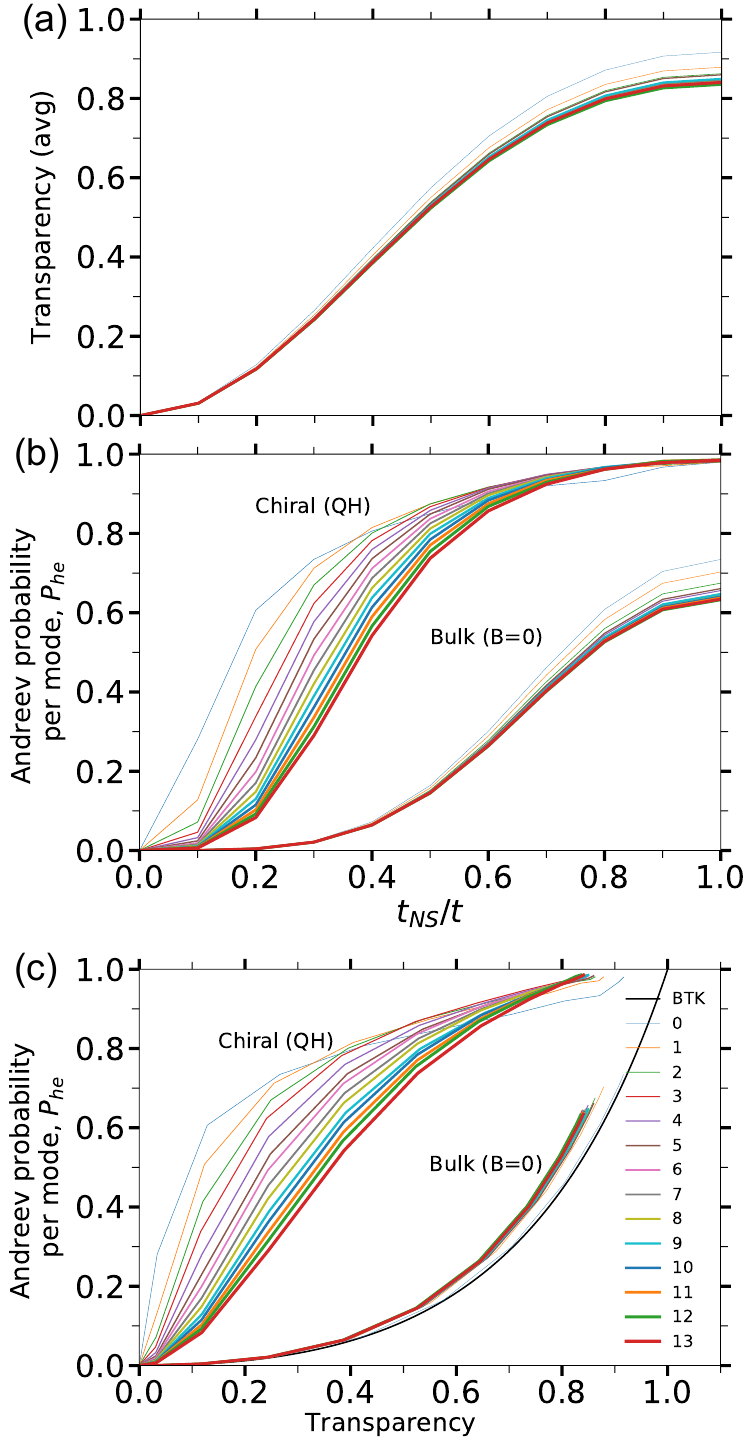}
\caption{Dependence of Andreev conversion on interface transparency, both at $B\!=\!0$ and in the QH regime, for different graphene doping (color-coded lines).   
(a) Transparency (two-terminal transmission probability through the interface with $B\!=\!\Delta\!=\!0$) as a function of the graphene-super\-con\-duc\-tor coupling, $t_{NS}/t$.  (b) Andreev probability per mode as graphene-superconductor coupling varies.  (c) Andreev probability per mode reexpressed through the transparency from (a); the black line is the BTK result \citep{BTK-PRB82} for comparison.  
Color labeled $n$ corresponds to graphene doping 
$\mu_\textrm{gr}/E_L \!=\! 0.05 \!+ \!0.0679\, n$, 
yielding a uniform sampling of 5-93\% of the LLL.  
(Zigzag parallelgram, sparse-stitch\-ed interface.) 
}
\label{fig:transparency}
\end{figure}

\subsection{Conductance as a function of\\ Interface Transparency\label{subsec:InterfaceTransparency}}

We start by evaluating the \emph{transparency}, $T$, of the honey\-comb-to-square-lattice interface in the absence of both superconductivity and a magnetic field.  This is done by finding the transmission per mode from the graphene to the superconductor using a simple two-ter\-min\-al nanoribbon geometry. 
For each value of $t_{NS}$ and $\mu_\textrm{gr}$, the transmission probability is averaged over a few nanoribbon widths $L$ 
\footnote{For the smallest $\mu_\textrm{gr}$, wider systems needed to be used.}. 
Results for $\mu_\textrm{gr}$ throughout the LLL and $t_{NS}/t \in [0,1]$ are presented in 
Fig.\,\ref{fig:transparency}(a). 
The maximum transparency for this sparse-stitched interface is $\sim\!85\%$, demonstrating a transparent interface despite the sudden, drastic change in the lattice.  As expected, as $t_{NS}$ decreases, the transparency decreases smoothly.  Note that there is very little dependence on $\mu_\textrm{gr}$, apart from the value nearest the Dirac point.  

Second, we find the probability of Andreev conversion both at $B\!=\!0$ and in the QH regime (for the same values of $\mu_\textrm{gr}$ and $t_{NS}$), as shown in Fig.\,\ref{fig:transparency}(b). For $B\!=\!0$, this is the Andreev reflection probability per mode in the same nanoribbon geometry used to find the transparency.  For the QH regime, we find the chiral Andreev conversion through the zigzag parallelogram nanostructure, thereby avoiding interference effects between the two edge modes. Finally, we eliminate $t_{NS}$ by combining panels (a) and (b), and show in panel (c) the Andreev conversion as a function of the transparency.  

The zero field result is familiar: the classic Blonder-Tinkham-Klapwijk (BTK) result, obtained in a simple one-dimensional model with parabolic dispersion and a $\delta$-function barrier, is $P_{he}\!=\!T^{2}/(2-T)^{2}$ \citep{BTK-PRB82}.
% ,BeenakkerPRB92,LesovikVoltageDependencePhysRevB.55.3146,Lesovik2011}. 
(See Ref.\,\cite{BeenakkerPRB92} for the generalization to the multimode situation.) 
%and finite bias variants of this are due to Beenakker and Lesovik, respectively). 
Fig.\,\ref{fig:transparency}(c) shows that for the entire range of $\mu_\textrm{gr}$ considered and despite the abrupt change in lattice and Fermi energy at the interface, the agreement between our tight-binding calculations and the BTK result (black line) is outstanding. 

In contrast, the result in the QH regime is strikingly different
\footnote{To compare chiral Andreev conversion and conventional Andreev reflection as in Figs.\,\ref{fig:transparency} and S4, one must start with the highest transparency interface, involving no intervalley scattering, as the reference point. 
Simply setting $t_{NS}/t\!=\!1$ does not necessarily achieve an ideally transparent interface since it does not eliminate scattering between the honeycomb and square lattices.  As discussed in Ref.\,\cite{AlexeyB_InfInter_PRB25}, it is necessary to also coarsely adjust $\mu_{S}$.}. 
First, the probability of Andreev conversion is significantly higher in the QH regime, especially at low transparency.  The larger Andreev probability is plausible from a semiclassical point of view: the incoming particle reflects many times from the interface before exiting (skipping orbits) while at $B\!=\!0$ in a clean system it reflects only once.  
Furthermore, the sign of the curvature of $P_{he}(T)$ is different in the two cases.  Thus the zero field result (positive curvature) is sensitive to transparency and becomes very small for $T\!\alt\! 0.3$.  In contrast, the negative curvature in the QH regime implies that chiral Andreev conversion is insensitive at moderate to high transparency, becoming sensitive only at low transparency and so becoming small only at very low transparency ($\alt\! 0.05$).  

The semiclassical argument in terms of skipping orbits suggests that this robustness is quite general, independent of disorder and details of the superconductor (such as its lattice structure).  The process of Andreev reflection itself is an interface property and so not very sensitive to disorder \cite{BeenakkerPRB92}.  
Disorder does, of course, cause multiple Andreev reflections at $B\!=\!0$ \cite{Beenakker-RMT-RMP97, LesovikSadovskyyRev11}, however not as systematically as for the skipping orbits produced by the magnetic field.  
On the other hand, disorder should have little effect in the quantum Hall regime (qualitatively) unless it is very strong (such that the Landau level structure is destroyed).  Likewise, the fact that graphene electrons experience repeated Andreev reflection at the interface is independent of details of the superconductor.  
Thus, as in the clean results shown here, we expect that chiral Andreev conversion is more robust in general.  

A second clear difference between the two regimes is that chiral Andreev conversion depends on the graphene filling, $\mu_\textrm{gr}/E_L$, while the $B\!=\!0$ Andreev reflection is completely independent of $\mu_\textrm{gr}$.  In fact, Andreev conversion is better for lower fillings: for small $\mu_\textrm{gr}$, $P_{he}$ is substantial for surprisingly small transparency.  This trend is consistent with the better electron-hole hybridization at the interface \cite{Phe-filling} seen in our previous work \cite{AlexeyB_InfInter_PRB25}.   

\subsection{Conductance as a function of\\ Chemical Potential\label{subsec:NanostrucTransport}}

The dependence of the chiral Andreev conversion on the chemical potential 
provides a striking illustration of the importance of intervalley scattering at corners.  Indeed, results for 
$P_{he}(\mu_\textrm{gr})$ were presented in Fig.\,\ref{fig:SummaryPA} as part of the summary in the introduction.  

First, consider an inverted trapezoid, which has \emph{zero} intervalley scattering corners (Sec.\,\ref{subsec:InvertTrap}).  In this case, $P_{he}\!=\!0$ independent of both $t_{NS}$ and $\mu_\textrm{gr}$. 

Next, consider a parallelogram nanostructure, which has \emph{one} intervalley scattering corner.  For a fully transparent interface, the output is entirely in the hole sector for all chemical potentials (Sec.\,\ref{sec:SimpleCases}).  However, for a partially transparent interface, we saw in Fig.\,\ref{fig:transparency}(b) some weak dependence on chemical potential.  Fig.\,\ref{fig:SummaryPA}(a) shows this directly as a function of $\mu_\textrm{gr}$.  For all values of $t_{NS}$ in the parallelogram, the dependence is smooth and monotonic 
\cite{Phe-filling}.  
$P_{he}(\mu_\textrm{gr})$ mostly varies slowly---there is rapid variation only for small $t_{NS}$.  

Finally, consider a trapezoid nanostructure, which has \emph{two} intervalley scattering corners (Sec.\,\ref{subsec:Trapezoid}).  Now both interface modes are populated at the first corner and recombined at the second, as in a Mach-Zender interferometer.  [The beating between the two modes was shown in Fig.\,\ref{fig:beating-trap}(b).]  $P_{he}(\mu_\textrm{gr})$ is oscillatory in this case, Fig.\,\ref{fig:SummaryPA}(b).  Note that for $t_{NS}\!=\!0.3$, the parallelogram result in panel (a) shows that a single intervalley corner acts as a beam splitter ($P_{he}\!\sim\! 0.5$).  We see in panel (b) that for such an intermediate value of $t_{NS}$ in the trapezoid, the oscillations are dramatic.  In an experiment, the magnitude would be reduced by any decoherence or loss processes that were present.

\section{Beyond Zigzag Edges \label{sec:BeyondZigzagEdges}}

Though the nature of the graphene edges is not controlled in the downstream resistance experiments (see e.g.\ \citep{LingfeiGlebCAESNatPhys20, SahuDas_Noise_PRB21, LingfeiGlebLossPRL23}), our results above are exclusively for zigzag edges because  
they are thought to be representative of an arbitrary termination of the honeycomb lattice  \citep{AkhBeenBoundaryPRB08, vanOstaayReconstructPRB11, WurmBaranger_Interfaces_NJP09, GiritZettl_GraphEdge_Science09, TkachovHentschel_PRB12}. In this section, we turn to considering edges that are \emph{not} ideal zigzag edges.  We start by briefly discussing the special case of armchair edges (Sec.\,\ref{subsec:Armchair-Rectangle}) and then move on to  nanostructures with arbitrary angles (Sec.\,\ref{subsec:ArbAngle-examp}), which leads to a more general characterization of intervalley-scattering versus valley-preserving corners (Sec.\,\ref{subsec:ArbAngle-general}).   Finally, we discuss in Sec.\,\ref{subsec:SmoothCorners} the experimentally important issue of corners that are rounded.  

\subsection{Rectangle: Role of Armchair Edges \label{subsec:Armchair-Rectangle}}

\begin{figure}[b]
\includegraphics[width=2.5in]{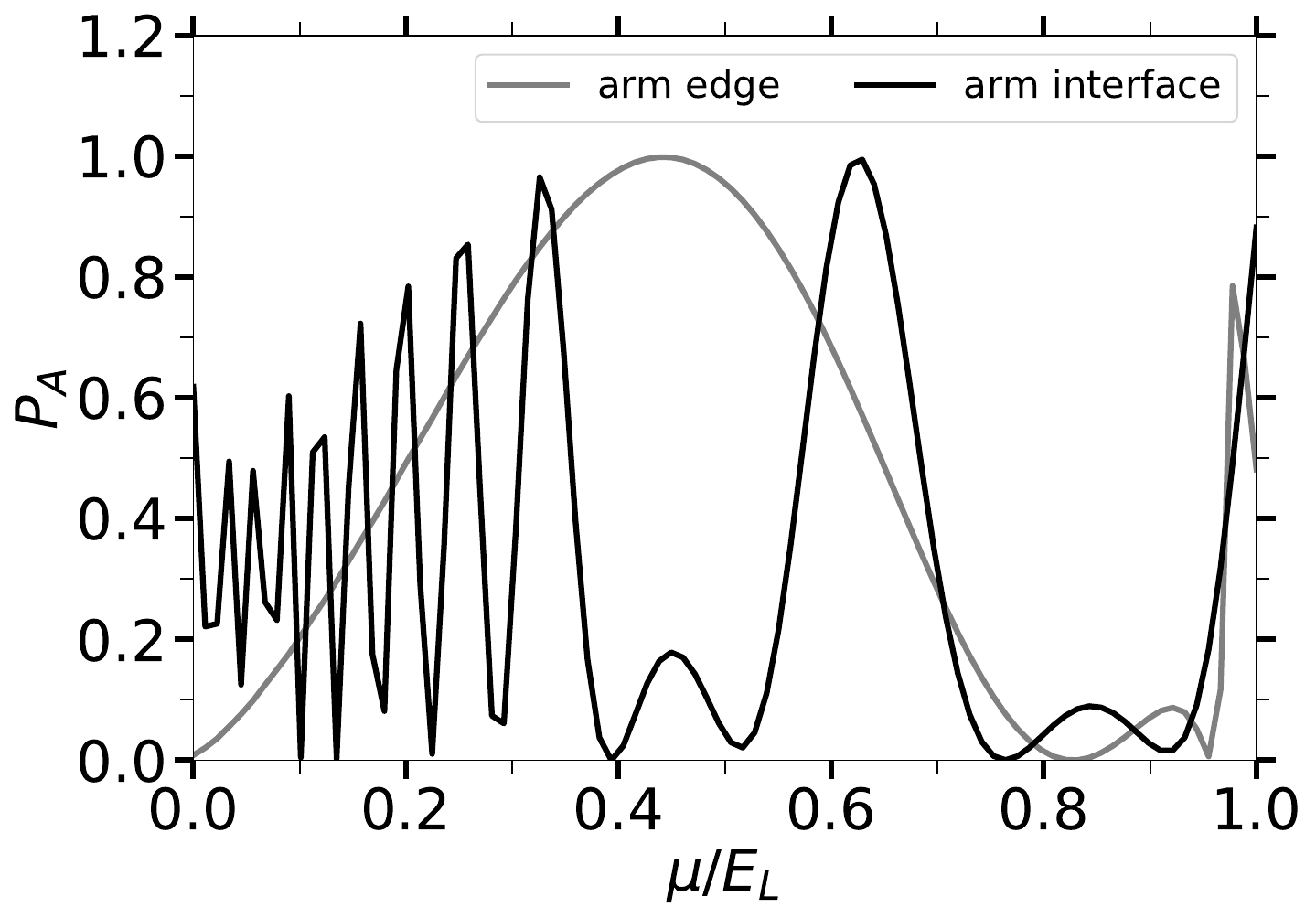}
\caption{Andreev conversion probability, $P_{he}$, as a function of the graphene doping $\mu_\textrm{gr}/E_{L}$ for 
a rectangular nanostructure (blue lines in Fig.\,\ref{fig:Geometry}).  $t_{NS}/t\!=\!1$ for both curves.  Oscillations caused by interference are present because the corners at the entrance and exit of the interface \emph{both} cause intervalley scattering. 
The rectangular case is special and atypical.  
(Standard parameters, see Sec.\,\ref{subsec:Parameters}.)
}
\label{fig:PA-armchair} 
\end{figure}

The armchair edge is a special case in which states in the $K$ and $K'$ valleys are mixed. This comes about because the Dirac points in the two-dimensional bulk Brillouin zone are both projected onto the center of the one-dimensional zone for the edge  states \citep{Abanin_QHGraph_PRL06,DelplaceMontambauxPRB10,BreyFertig_EdgePRB06}. 
The LLL armchair edge states, then, have weight on both the A and B sublattices. Since armchair edges are perpendicular to zigzag edges, replacing the parallelogram or trapezoid of the last section by a rectangular nanostructure [$\theta_{1}\!=\!\theta_{2}\!=\!\pi/2$ in the geometry of Fig.\,\ref{fig:Geometry} (blue lines)] necessarily introduces armchair edges.  

First, consider the case when the interface with the superconductor is a zigzag edge, as before, and the incoming and outgoing (vertical) edges are armchair. The incoming electron has weight in both valleys, and so both hybrid interface modes are excited. They are then mixed at the junction between the interface and the outgoing armchair edge. 
Interference effects thus occur, and, indeed, we see in Fig.\,\ref{fig:PA-armchair} that $P_{he}$ depends on the chemical potential in the graphene. However, the variation is rather slow because for a fully transparent interface the two modes are nearly degenerate for all $\mu_\textrm{gr}$, and so only a small change in phase difference can accumulate 
while varying $\mu_\textrm{gr}$  
($k_{1}-K\equiv\delta k\approx\delta k'\equiv k_{2}-K'$).  A partially transparent interface yields a larger splitting between the hybrid modes along the interface, and thus more striking interference effects. 

Now consider the opposite type of rectangle: the interface is armchair and the incoming and outgoing graphene edges are zigzag. The incoming electron is in the valley of the propagating mode.  It excites both modes along the interface since each has weight in both $K$ and $K'$.  
In contrast to the previous case, the two hybrid interface modes can have a large splitting 
($\delta k \sim\!1/l_B$) 
even at full transparency \cite{LingfeiGlebCAESNatPhys20}.  Thus, a large variable phase difference between the two hybrid modes accumulates along the interface, and the resulting interference makes 
$P_{he}(\mu_\textrm{gr})$ oscillate rapidly, as shown in Fig.\,\ref{fig:PA-armchair}.  Though interference effects are likely to be stronger in systems with armchair edges, we emphasize that such nanostructures are rare \cite{AkhBeenBoundaryPRB08, vanOstaayReconstructPRB11, Manesco-QHgrS_SP22}. 

\begin{figure}
\includegraphics[width=3.2in]{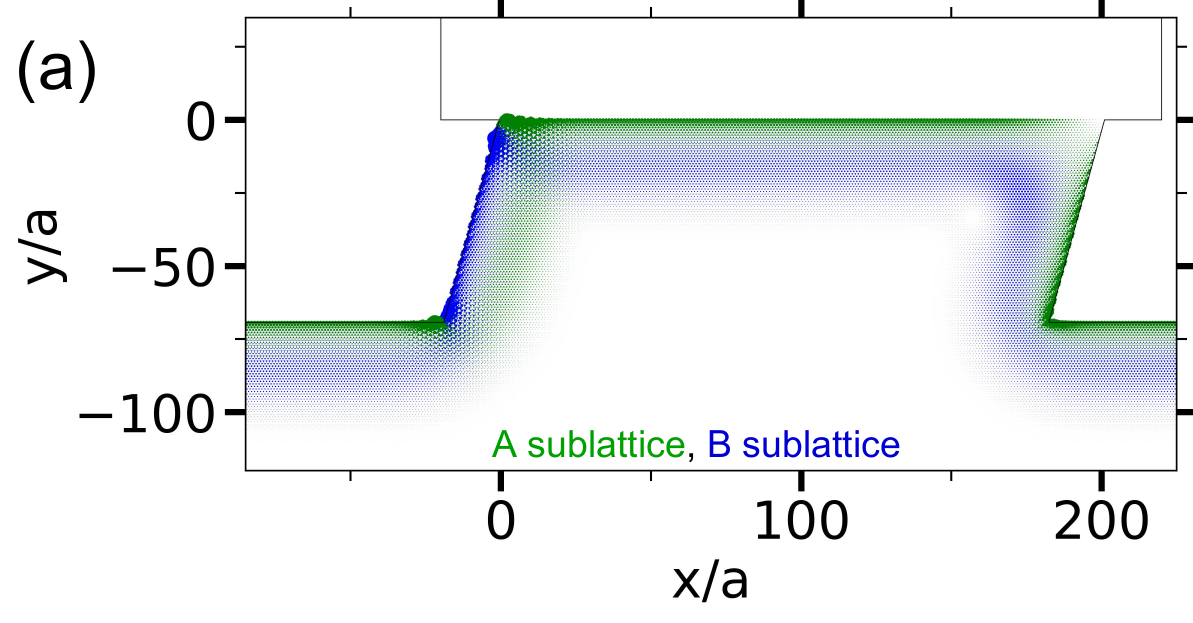}
\includegraphics[width=3.2in]{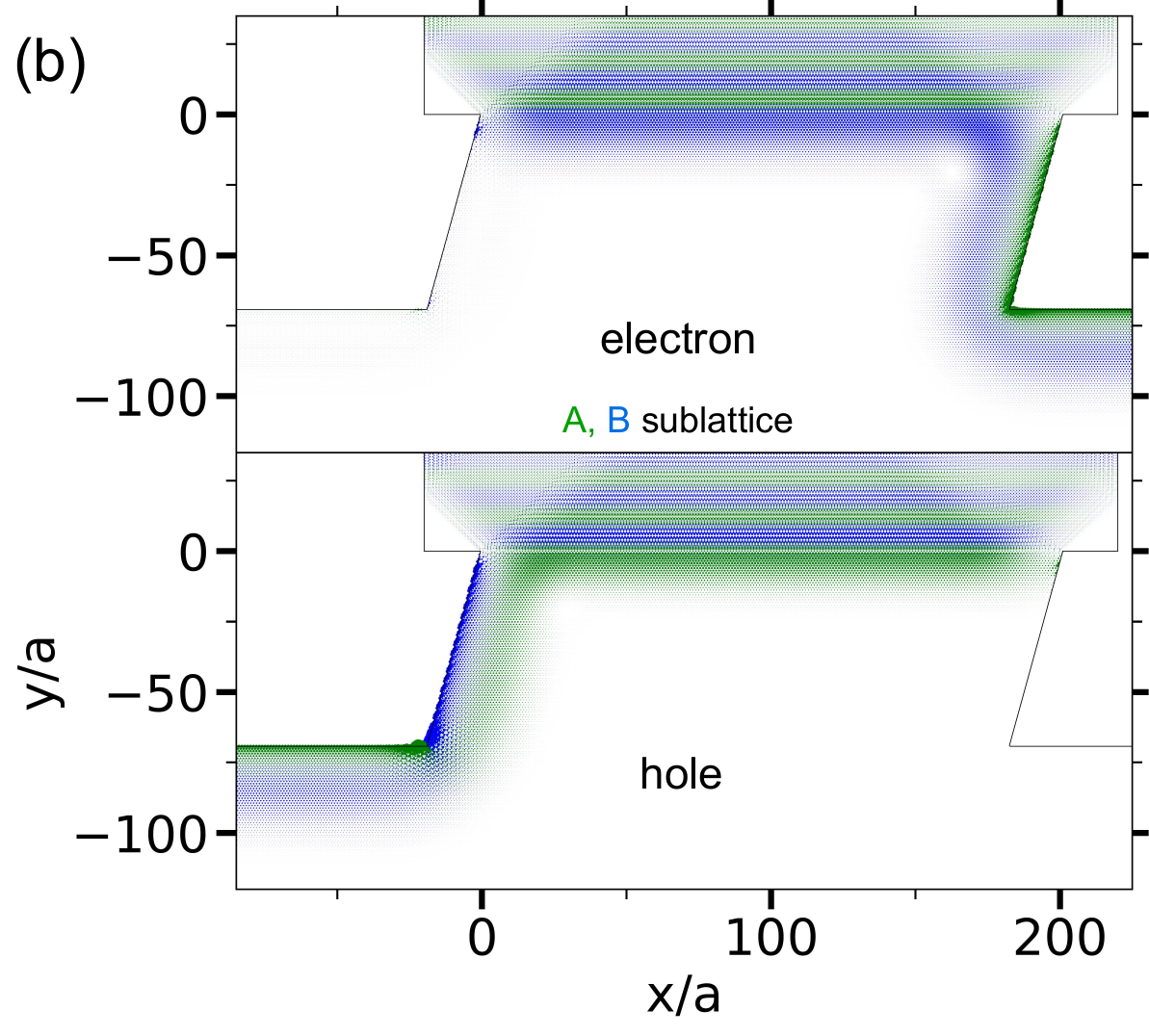}
\caption{Scattering state, $|\psi^{\beta}(x,y)|^2$ with $\beta\!=\!e,h$, for a parallelogram 
with $\theta_1\!=\!\theta_2\!=\!75 ^\circ$ at (a) zero or (b) full coupling ($t_{NS}/t\!=\!0$ or $1$).   
%In each panel the electron (hole) component is in the upper (lower) part \cite{PlottingMethod}.  
These are analogous to the results for the zigzag parallelogram in Figs.\,\ref{fig:DisconScatState} and \ref{fig:ScatterState_parallelogram} and show the same features.  (Standard parameters, see Sec.\,\ref{subsec:Parameters}.) 
}
\label{fig:ArbAngle}
\end{figure}

\subsection{Arbitrary Angle Nanostructures: Examples \label{subsec:ArbAngle-examp}}

For a parallelogram with $\theta_1\!=\!\theta_2\!=\!75^\circ$, 
Fig.\,\ref{fig:ArbAngle} shows the scattering state for both an opaque and a high-transparency interface. 
In comparison to the results for the fully zigzag parallelogram in Figs.\,\ref{fig:DisconScatState} and \ref{fig:ScatterState_parallelogram}(a), the wavefunctions here show more irregularity, but the overall results are the same.  
For the high-transparency interface, the particle stays in the $K$ valley through the first two corners, connects smoothly to the propagating hole state at the third corner, and then scatters at the fourth corner in order to propagate as a hole along the outgoing edge. Andreev conversion is large---$P_{he}$ is close to unity and relatively independent of parameters. 

\begin{figure}
\includegraphics[width=2.5in]{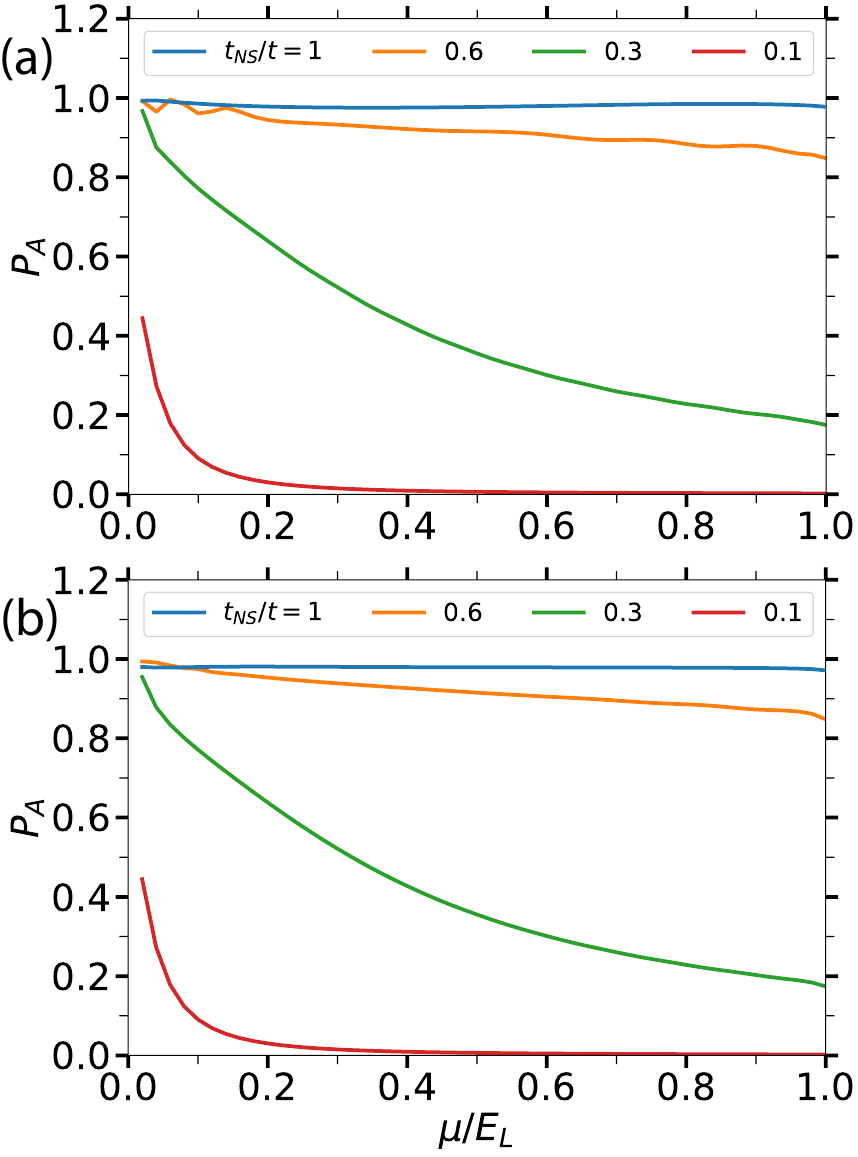}
\caption{Andreev conversion probability $P_{he}$ as a function of the graphene doping $\mu_\textrm{gr}/E_{L}$ for two non-zigzag nanostructures: (a) $\theta_1\!=\!\theta_2\!=\!75^\circ$ and (b) $\theta_1\!=\!45 ^\circ$, $\theta_2\!=\!75 ^\circ$.  
Line color codes the strength of graphene-superconductor coupling, $t_{NS}/t$.
%The strength of graphene-superconductor coupling, $t_{NS}/t$, is indicated by the line color.  
For full transparency ($t_{NS}/t\!=\!1$, blue), Andreev conversion is nearly perfect for both structures, independent of $\mu_\textrm{gr}$.  
Results are qualitatively similar to those for the zigzag parallelogram in Fig.\,\ref{fig:SummaryPA}(a).  (Standard parameters in Sec.\,\ref{subsec:Parameters}.)
}
\label{fig:ArbAnglePhe}
\end{figure}

The Andreev conversion probability in the LLL for two non-zigzag nanostructures
%, related to their conductance by $G\!=\!(e^2/h) \left[1-2P_{he}\right]$, 
is shown in Fig.\,\ref{fig:ArbAnglePhe}.  The first case is the parallelogram $\theta_1\!=\!\theta_2\!=\!75 ^\circ$, as for Fig.\,\ref{fig:ArbAngle}.  
The second structure is a quadrilateral with one angle larger than $60^\circ$ and one less, $\theta_1\!=\!45 ^\circ$ and $\theta_2\!=\!75 ^\circ$.  Comparing to $P_{he}(\mu_\textrm{gr})$ for the all zigzag structure in Fig.\,\ref{fig:SummaryPA}(a), we see that 
%for a partially transparent interface 
there is more variation here but that the quantitative differences are not large.  
Qualitatively, the results for all three structures are the same: a slowly varying probability of Andreev conversion which is larger for lower filling.

\subsection{Arbitrary Angle Nanostructures: General Argument \label{subsec:ArbAngle-general}}

In discussing results for structures with zigzag edges (Secs.\,\ref{sec:SimpleCases}-\ref{sec:Conductance}), we emphasized that the valleys of the propagating state on the incoming and outgoing edges are critical in determining the net Andreev conversion.  For an edge at an arbitrary angle,  
it is likewise critical to know the valley properties of the propagating state.  This can be deduced from sublattice-valley locking and the boundary condition for an edge at an arbitrary angle. 

First, the boundary condition at an arbitrary edge has been shown to be essentially that of a zigzag edge in the large majority of cases \citep{AkhBeenBoundaryPRB08,vanOstaayReconstructPRB11}. 
For instance, consider two half sheets of graphene, one with a zigzag edge for which the boundary condition is $\psi_{B}\!=\!0$ (where $\psi_{B}$ refers to the amplitude on the B sublattice), and a second with an arbitrary edge at angle $\theta$ with respect to this zigzag edge.   Ref.\,\citep{AkhBeenBoundaryPRB08} shows that $\psi_{B}\!=\!0$ is the appropriate boundary condition for any edge with $\left|\theta\right|<30^\circ$ for states in the lowest energy band 
\footnote{Compared to the zigzag case, the unit cell for the edge at angle $\theta$ is larger (possibly much larger) and the Brillouin zone is correspondingly smaller, with the graphene conduction band folded back into the reduced zone causing multiple bands.  
The demonstration in Ref.\,\citep{AkhBeenBoundaryPRB08} has an additional technical requirement of a ``minimal boundary'' that we neglect here.}. 
The edges with $\theta\!=\!\pm30^\circ$ are armchair edges---a special case (Sec.\,\ref{subsec:Armchair-Rectangle}).  Edges with $\left|\theta\right|>30^\circ$ are, of course, closer to a different zigzag edge, and that edge should then be used to determine the appropriate boundary condition.  

Second, sublattice-valley locking (valid in the LLL) then implies that a boundary condition on a single sublattice affects states in only one valley.  
For example, if the bulk LLL states in valley $K$ are on the B sublattice, then the boundary condition $\psi_{B}\!=\!0$ implies that the LLL chiral edge state is in the $K$ valley. 
Since the boundary condition is of zigzag type for most edges, the chiral edge state for an edge at angle $\theta$ will be in the same valley as that for the closest zigzag edge 
\footnote{For an arbitrary edge, ``valley'' refers to the Dirac points folded back into the reduced Brillouin zone.}. 

The sublattice structure of the zigzag edges alternates as $\theta$ increases: in our example, the zigzag edges at $\theta\!=\!\pm60^\circ$ will have the boundary condition $\psi_{A}\!=\!0$, and hence the chiral edge states will be in the $K'$ valley. We find, then, that for $30^\circ<\theta<90^\circ$, the valley of the chiral edge state will be opposite to that of the original zigzag edge. Note that one can have a small change in edge direction, say from $\theta\!=\!25^\circ$ to $\theta\!=\!35^\circ$, and still have a change in the valley of the edge state, thus requiring intervalley scattering to connect the two edge states \citep{WurmBaranger_Interfaces_NJP09}.  

The classification of corners as either \emph{intervalley-scattering} or \emph{valley-preserving} introduced in Sec.\,\ref{sec:SimpleCases} can thus be readily extended to arbitrary angles.  From the argument in Ref.\,\citep{AkhBeenBoundaryPRB08} for the boundary condition at a non-zigzag edge, we see that the key factor is whether the majority of terminal lattice sites belong to the A or B sublattice.  
A corner that connects two edges in which the majority sublattice is different (the same) is intervalley-scattering (valley-preserving).  While creating a valley-mixing armchair edge requires precise angular control of the edge \citep{Manesco-QHgrS_SP22}, creating intervalley-scattering or valley-preserving corners does not, allowing room for creating such corners in future experimental design of atomistically controlled graphene nanostructures.

From this simple rule for the valley structure of LLL edge states at an arbitrary angle, the Andreev conversion properties follow as for the zigzag edges in Sec.\,\ref{sec:ZigzagScattering}. 
For a fully transparent graphene-superconductor interface, either the electron or hole exits ($P_{he}\!=\!0$ or $1$), depending on which matches the valley of the propagating state on the outgoing edge (recall that the propagating electron and hole are in opposite valleys).  This is Eq.\,(\ref{eq:Phe-ideal-AkhBeen}). 
For a partially transparent interface, however, intervalley scattering may be needed to make the connection. Our numerical results for non-zigzag structures (Figs.\,\ref{fig:ArbAngle} and \ref{fig:ArbAnglePhe}) indeed follow these guidelines.

\subsection{Smooth Corners \label{subsec:SmoothCorners} }

Sharp corners clearly play a critical role in the properties of a nanostructure, in particular in the amount of Andreev conversion that occurs. In many experiments, however, the corners are unlikely to be atomically sharp, even if the interface and superconductor are largely free of disorder. It is natural, then, to consider structures with smooth (rounded 
\footnote{Let us specify more precisely what we mean by a smooth or rounded corner. At the junction between a QH edge state and the hybrid mode along the interface, it is likely that evanescent states will be needed in order to successfully match the states. Suppose the least evanescent of these states decays on a length scale $d$. Then the radius of curvature of the rounded corner should be at least a few times bigger than $d$.})  
corners at both ends of the graphene-superconductor interface and to ask whether this study can shed light on their properties.  

First, consider structures in which the superconductor contacts only the straight edge between the two rounded corners---the smooth corners occur entirely within the graphene nanostructure with QH edge states.  Results for this case can be immediately deduced from the sharp corner results.  
Because there is no corner at the junction between the QH edge state and the chiral Andreev interface modes, the valley of the propagating state stays the same from the incoming edge to the outgoing edge.  Thus, $P_{he}\!=\!0$ for full transparency, as per Eq.\,(\ref{eq:Phe-ideal-AkhBeen}).  Because there are no intervalley corners along the interface  
\footnote{There may, of course, be intervalley scattering at some point along the smooth corner---for instance, if an interval of armchair edge is encountered. Such intervalley scattering away from the interface with the superconductor does not affect the Andreev conversion. Rather it simply modifies the valley of the incoming or outgoing QH edge state. What is important is the edge state immediately before the interface, not the one prior to the corner that allows the QH edge state to interface with the superconductor. }, 
even for partial transparency $P_{he}\!=\!0$.  We find, then, that for all smooth connections to a straight interface, Andreev conversion is completely suppressed.  

Now consider structures in which the superconductor does contact the graphene for part of the rounded corner.  There are two key considerations: (i)~the nature of the QH edge at the junction with the hybrid interface mode and (ii)~whether in the absence of superconductivity (i.e.\ with $t_{NS}\!=\!0$) there would be intervalley scattering in the interface interval. 
% (or within a superconducting correlation length of the interface). 

At full transparency, only consideration (i) is active: Andreev conversion will follow Eq.\,(\ref{eq:Phe-ideal-AkhBeen}), namely that $P_{he}\!=\!0$ if the incoming and outgoing edge are of the same type and $P_{he}\!=\!1$ if they are of opposite types. (The type of edge here refers to the zigzag edge to which it is closest.)  

For an interface with only partial transparency, consideration (ii) comes into play as well.  
If the geometry is such that a locally armchair region occurs along
the interface with the superconductor, then intervalley scattering will occur---despite it being rounded, this is an intervalley-scattering corner producing mode mixing.  There will be partial Andreev conversion, $0<P_{he}<1$, that does not follow Eq.\,(\ref{eq:Phe-ideal-AkhBeen}).  
The qualitative behavior is the same as for sharp corners: for one intervalley corner $P_{he}$ varies slowly with parameters (parallelogram, Sec.\,\ref{subsec:Parallelogram}), while for two (or more) intervalley corners, interference occurs leading to 
oscillation of $P_{he}$ (trapezoid, Sec.\,\ref{subsec:Trapezoid}). 

We note that the results of \citep{JoRoulleau_ValleySplitter_PRL21} suggest that it may be possible to use electrostatic gates to move the intervalley-scattering corner in and out of the proximity region and thereby switch between different Andreev conversion conductance patterns.  While smooth potentials from gates do not typically cause intervalley scattering in bulk graphene,
%(and can be treated by a Dirac equation), 
this is not true, independently of smoothness, for generic terminated-lattice nanostructures  \citep{Akhmerov_ValleyValve_PRB08, RainisFazioNanoribPRB09, TrifunovicBrouwer_GraphPN_PRB19}, and so intervalley scattering should remain for electrostatically smooth corners.

\section{Conclusion \label{sec:Conclusion}}

\subsection{Summary}

We have used scattering states to study Andreev conversion in QH graphene nanostructures connected to a superconductor.  The focus is on the LLL and clean nanostructures (no disorder) with sharp corners.  This work is based on the assumption (see Sec.\,\ref{subsec:BdG-Approxes} and \cite{AlexeyB_InfInter_PRB25}) that the one-body potential, magnetic field, and pairing potential change abruptly at the interface. 

Most of our results are obtained on structures with zigzag edges because these represent the properties of typical edges 
\cite{AkhBeenBoundaryPRB08, vanOstaayReconstructPRB11, Manesco-QHgrS_SP22}; however, some results for non-zigzag edges are presented in Sec.\,\ref{subsec:ArbAngle-examp}.  
An argument for structures with sharp corners between straight edges at arbitrary angles is developed in Sec.\,\ref{subsec:ArbAngle-general}, which then leads to general statements in Sec.\,\ref{subsec:SmoothCorners} about structures with corners that are smooth or rounded.  

For a well-matched, transparent interface (an ideal connection between the graphene and the superconductor), the connection of the scattering state between the interface and both the incoming and outgoing graphene edges is \emph{completely smooth}. 
There is no intervalley scat\-tering---the valley of the excitation is preserved across the connection---and as a result, the Andreev conversion, $P_{he}$, is either zero or one.  We found this for both types of square lattice superconductors (dense and sparse stitching, Fig.\,\ref{fig:Stitching}). Thus the simple Akhmerov-Beenakker result Eq.\,(\ref{eq:Phe-ideal-AkhBeen}) \citep{AkhmerovValleyPolarPRL07} is obeyed in this ideal limit.  

A partial-transparency interface is quite different. First, the average $P_{he}$ varies smoothly between zero and one as a function of transparency.  
Compared to our $B\!=\!0$ results, chiral Andreev conversion is larger, more robust to reduced transparency, and sensitive to the graphene chemical potential.  While the $B\!=\!0$ results follow the classic BTK expression, the QH results do not.  Based on semiclassical skipping orbit arguments, we expect that chiral Andreev conversion will be more robust in general. 

Intervalley scattering can occur at the entrance or exit corner of a partial-transparency interface. Typically, such scattering occurs if the corner is an inter\-valley-scat\-ter\-ing corner 
% in the corresponding terminated lattice structure (i.e., with $t_{NS}=0$) 
when $t_{NS}\!=\!0$,
and it causes both electron-hole modes to appear in the scattering state.  Furthermore, intervalley-scattering corners couple more strongly to the superconductor, leading to Andreev intervalley scattering.  Compared to expectations based on the $e$-$h$ hybridization of the interface modes, the resulting Andreev conversion is larger. 

If there is one intervalley-scattering location in the structure, $P_{he}$ changes monotonically as a function of chemical potential. There are no interference effects, but Eq.\,(\ref{eq:Phe-ideal-AkhBeen}) does not apply. 

Interference effects occur naturally in $P_{he}$ when there is intervalley scattering at \emph{both} the entrance and exit to a partial-transparency interface. 
Scattering at the entrance populates both hybrid modes, and these amplitudes are recombined by the exit scattering, forming a kind of Mach-Zehnder interferometer. 
The partial transparency of the interface provides \emph{two} physical effects, both of which must be present to produce interference: (i) intervalley scattering \emph{mixes} the two interface modes, and (ii) these two modes are \emph{not} valley degenerate (so that there is a difference in phase as they propagate along the interface). 

\subsection{Relevance to Experiments}
\label{subsec:Expts}

We close with some comments on the relevance of these results for experiments (see e.g.\ Refs.\,\citep{LingfeiGlebLossPRL23, LeeHarvardNatPhy17, LingfeiGlebCAESNatPhys20, SahuDas_Noise_PRB21, HatefipourShabani_QH-S_NL22}).  We make four main simplifications in our tight-binding treatment, as discussed in Secs.\,\ref{subsec:BdG-Approxes} and \ref{subsec:TightBind}.  The magnitude of Andreev conversion is considerably larger here than in the experiments because we treat a completely coherent, lossless model.  Since decoherence and loss are expected to be insensitive to nanostructure geometry or parameters such as $\mu_\textrm{gr}$, the trends and general conclusions that we find should be valid despite the presence of some loss and decoherence.  

Second, our highly simplified treatment of the magnetic field (an abrupt jump to zero in the superconductor) means that any vortex physics seen experimentally is, of course, lost.  Nevertheless, this simplified treatment has been extensively used in the past to successfully describe general features of superconductor-nor\-mal systems, and we expect the same to be the case here.  

We leave the issue of spin physics, 
% (the Zeeman effect and spin-orbit coupling), 
which is evident in the experiments, to future work.  In the absence of spin-orbit coupling, the spin projected results here can be used for each spin species: our previous work on infinite interfaces \cite{AlexeyB_InfInter_PRB25}, for instance, showed that changes introduced by the Zeeman effect are small when the interface is transparent.  
However, spin-orbit coupling (coming, presumably, from a heavy element in the superconductor) will mix the two spin channels in the LLL, leading possibly to additional effects. 

We emphasize, finally, that no disorder is included in the present results: we have treated completely clean, lattice matched systems.  In the experiments to date  (see e.g.\ Refs.\,\citep{LingfeiGlebLossPRL23, LeeHarvardNatPhy17, LingfeiGlebCAESNatPhys20, SahuDas_Noise_PRB21, HatefipourShabani_QH-S_NL22}), an alloy is used for the superconductor 
%(which is therefore type II) 
and so disorder is an essential part of the system.  
Atomic-scale disorder causes intervalley scattering which, then, nullifies arguments based on the valley of the edge state.  Disorder makes possible, for instance, pairing correlations between the propagating electron and hole on the terminated-lattice QH edge, and so makes Andreev conversion possible in the tunneling limit \citep{KurilovichDisorderAES_NCom23, Manesco-QHgrS_SP22}.  
% Other weaknesses of our study include a number of approximations in treating the interface, discussed in Sec.\,\ref{subsec:BdG-Approxes}. 

Despite the presence of disorder, scattering due to the geometry of the nanostructure, which we emphasize here, may still be important.  Such scattering can, for instance, change the electron-hole hybridization so that the output does not reflect the average electron-hole hybridization along the interface.  
Since the geometric scattering is largely independent of disorder, the trends as a function of parameters that we describe should generally hold.  One important difference in the phenomenology is that the interference pattern in the presence of disorder will be irregular \cite{LingfeiGlebCAESNatPhys20, Manesco-QHgrS_SP22, KurilovichDisorderAES_NCom23}, as observed experimentally \cite{LeeHarvardNatPhy17, LingfeiGlebCAESNatPhys20, SahuDas_Noise_PRB21} and as expected for a mesoscopic interference effect \cite{Beenakker-RMT-RMP97, LesovikSadovskyyRev11}, rather than the regular pattern found here (Fig.\,\ref{fig:SummaryPA}). 

On the other hand, superconductivity in pristine transition metal dichalcogenides has recently come under intense scrutiny \citep{RossiTriola_vdW-SO_AnnPhys20, Manzeli_TMDreview_NRM17, Wang_ElectronicsTMD_NNano12}.  In fact, coexistence and coupling of QH graphene with superconducting 2D niobium diselenide has been demonstrated \citep{SahuBangalorePRL18}.  
It seems quite possible that transparent, low disorder interfaces between these materials and graphene will become available.  Our results will apply directly to these future systems. 
%Furthermore, results for the clean system provide a basis for ongoing studies of the disordered case. 

It is clear in any case that the appealing, geometric result Eq.\,(\ref{eq:Phe-ideal-AkhBeen}) has a limited range of applicability.  
%For a disordered system, the underlying assumptions about valleys are, of course, not valid.  But 
Even in the completely clean case, we have seen that it applies only if the graphene-superconductor interface is highly transparent such that intervalley scattering is absent.  Once the interface is not perfectly transparent, the two hybrid modes at the interface are not degenerate, and scattering at the corners, even if they are rounded, mixes the two modes causing interference effects.\\ 

\begin{acknowledgments}
We thank Ethan G. Arnault, Anthony David, Gleb Fin\-kel\-stein, Gu Zhang, and Lingfei Zhao for helpful discussions. 
This work was supported in part by the U.S.\ Department of Energy, Office of Basic Energy Sciences, Materials Sciences and Engineering Division, under Award No.\ DE-SC0005237.
A.B.\ acknowledges support by the National Defense Science and Engineering Graduate (NDSEG) Fellowship Program. 
\end{acknowledgments}

\smallskip
The data that support the findings of this article are openly available \cite{Data_GrS-Nanostruc}.

\bibliography{QTransport_2025-11,\jobname} 

\begin{thebibliography}{100}
\bibitem [{Plo()}]{PlottingMethod}%
  \BibitemOpen
  \href@noop {} {}\bibinfo {note} {At each lattice site $i$, we plot a disk
  with area proportional to $|\psi^\beta_i|^2$ ($\beta\!=\!e,h$); the disk is
  green for sublattice A sites and blue for B. The electron (hole) component is
  in the upper (lower) panel. For the sparse lattice, the graphene sublattice
  structure does not continue into the square lattice and so the disks are
  colored orange. The intensity scale is chosen such that (for our standard
  parameters) the disks on the terminal A sites of the incoming electron state
  just barely overlap. Disks are smaller and non-overlapping at higher doping
  $\mu_\textrm{gr}/E_L\!=\!1/2$ since the surface state is less localized to
  the boundary \citep{BreyFertig_EdgePRB06}. Note that there is saturation in
  the wavefunction peaks at the intervalley-scattering corners.}\BibitemShut
  {Stop}%
\bibitem [{Phe()}]{Phe-filling}%
  \BibitemOpen
  \href@noop {} {}\bibinfo {note} {More Andreev conversion at lower graphene
  filling [monotonic decrease of $P_{he}(\mu_\textrm{gr})$] can be understood
  through the spectrum of the QH edge and interface modes. In
  \cite{AlexeyB_InfInter_PRB25} we pointed out that $e$-$h$ hybridization is
  connected to a crossing in the BdG spectrum of QH edge states at
  $E\!=\!\mu_\textrm{gr}$ (terminated-lattice, no superconductivity). Small
  $t_{NS}$ converts this to an anti-crossing, and then larger $t_{NS}$ causes
  full hybridization and the valley-degenerate spectrum (see Figs.\,6-9 in
  \cite{AlexeyB_InfInter_PRB25}). Smaller $\mu_\textrm{gr}$ means that the
  initial crossing is closer to the chemical potential ($E\!=\!0$), and thus
  $e$-$h$ hybridization is enhanced.}\BibitemShut {Stop}%
\end{thebibliography}

%\newpage
%\bibliographystyle{unsrt}
%\phantomsection\addcontentsline{toc}{section}{\refname}\bibliography{../../QTransport_2024-10,refs_final}

%%% SUPPLEMENTAL MATERIAL %%%%%%%%%%%%%%%%%%%%%%%
%\widetext
\clearpage
%\newpage
%\large

\setcounter{equation}{0}
\setcounter{section}{0}
\setcounter{page}{1}
\newcounter{Sfigure}

\global\long\def\theequation{S\arabic{equation}}
\global\long\def\thefigure{{S\arabic{Sfigure}}}
\global\long\def\thesection{S\arabic{section}}

\onecolumngrid
\begin{center}
\textbf{{\large Supplemental Material for \\``Quantum {Hall} {Andreev} Conversion in Graphene Nanostructures''}}\\[0.1in]
Alexey Bondarev, William H. Klein, and Harold U. Baranger\\[0.05in]
\emph{Department of Physics, Duke University, Durham, NC 27708-0305, USA}\\[0.05in]
%\today%% should be hard-wired to date of submission to arXiv
(1 July 2025)
\end{center}
\twocolumngrid

In this Supplemental Material, we collect results using the superconductor lattice not considered in the main text---dense or sparse stitching (shown in Fig.\,\ref{fig:Stitching} of the main text) as appropriate.  The energy spectrum, scattering states, and Andreev conversion probability, $P_{he}$, are shown for sparse stitching, and the dependence of Andreev conversion on interface transparency is given for dense stitching.  The results are very similar to those in the main text.  Thus, they provide a second example, demonstrating explicitly that the our conclusions do not depend on any particular stitching at the interface.

\begin{figure}[b]
\includegraphics[width=3.4in]{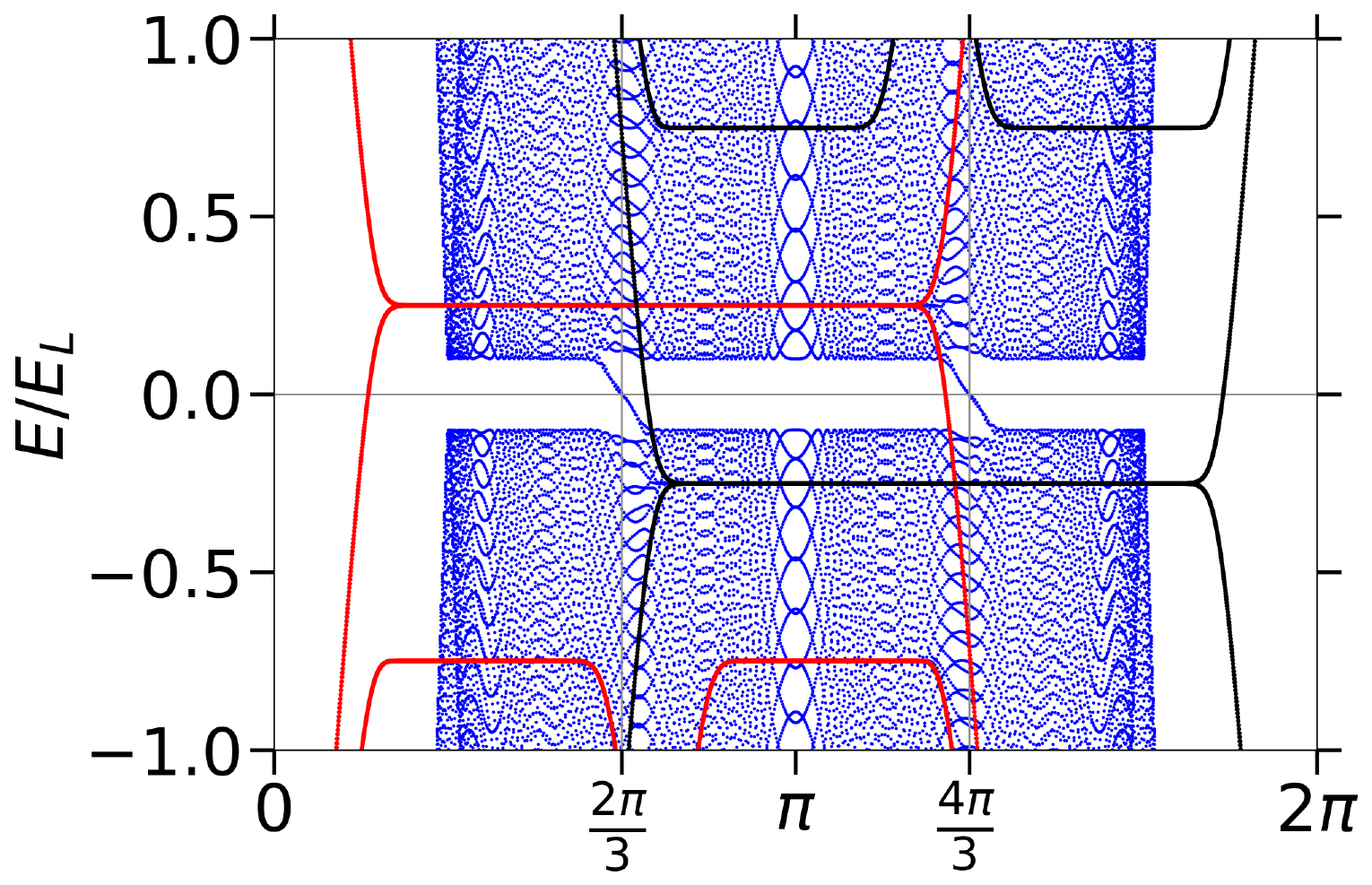}
\caption{The energy spectrum for QH states in a zig\-zag graphene nanoribbon [electron (black) and hole (red), $t_{NS}\!=\!0$] superposed with those for a transparent graphene-super\-conductor interface (blue, $t_{NS}/t\!=\!1$) in the \emph{sparse-stitched} case.  The chiral Andreev interface modes  have energies within the superconducting gap and have $E\!=\!0$ at $\!k=\!K$ and $K'$ (marked by vertical lines).  Compare to dense-stitching results in Fig.\,\ref{fig:Dispersion}.  (Parameters otherwise standard, see Sec.\,\ref{subsec:Parameters}.)
}
\label{Fig-S1_spectrum-sparse}
\end{figure}

First, in Fig.\,\ref{Fig-S1_spectrum-sparse} we show the energy spectrum of both the QH terminated lattice edge states and the interface states for an infinite, periodic sparse-stitched structure.  The states of the superconducting continuum do not cover the full Brillouin zone, in contrast to the dense-stitching case, because the lattice constant of the square and honeycomb lattices are the same (no zone folding).  As in the dense-stitched case, for this well-matched, transparent interface, the chiral Andreev interface modes go through the center of the valley.  

\stepcounter{Sfigure}

\stepcounter{Sfigure}\begin{figure}[b]    
\includegraphics[width=3.4in]{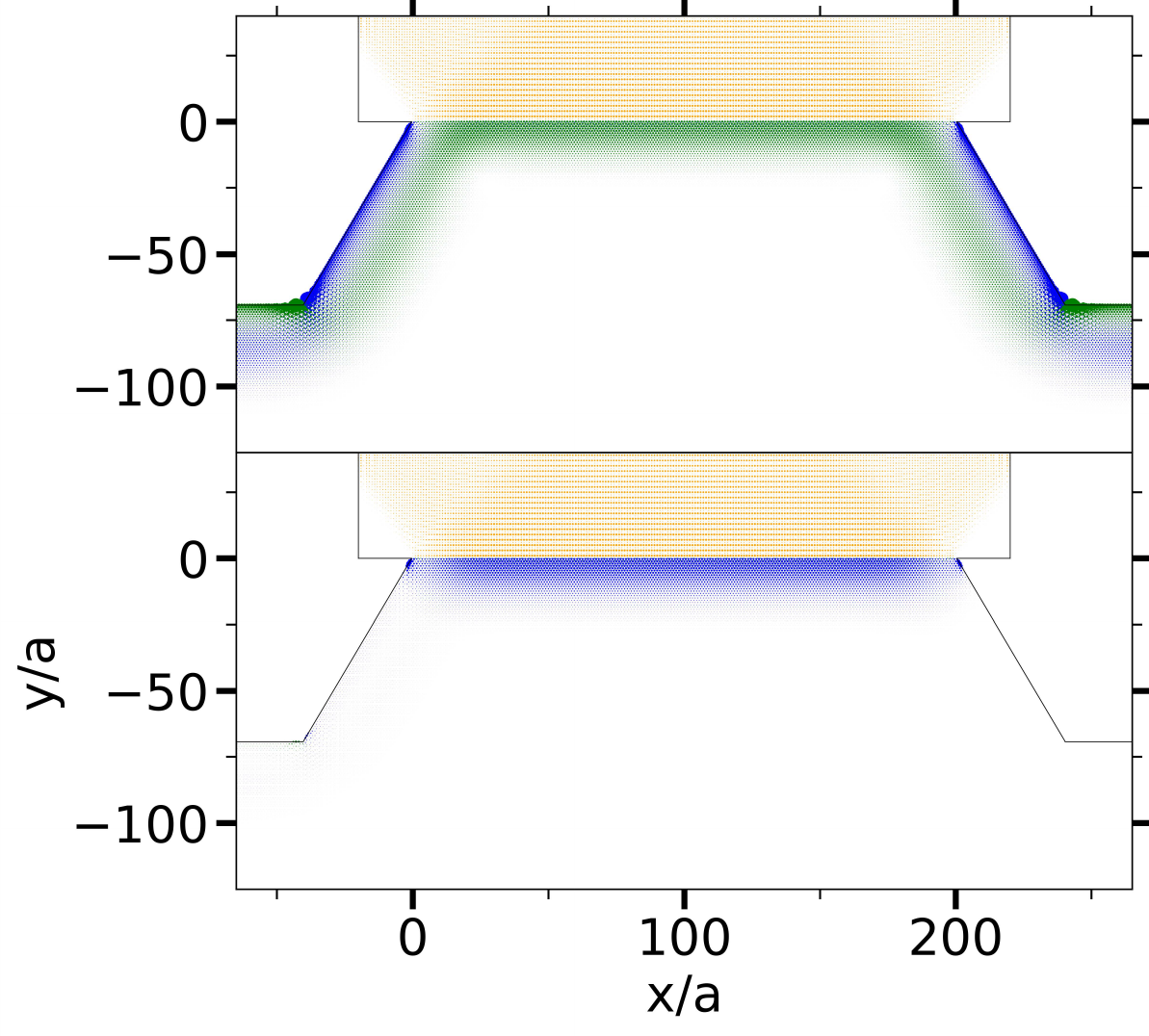}
\includegraphics[width=3.4in]{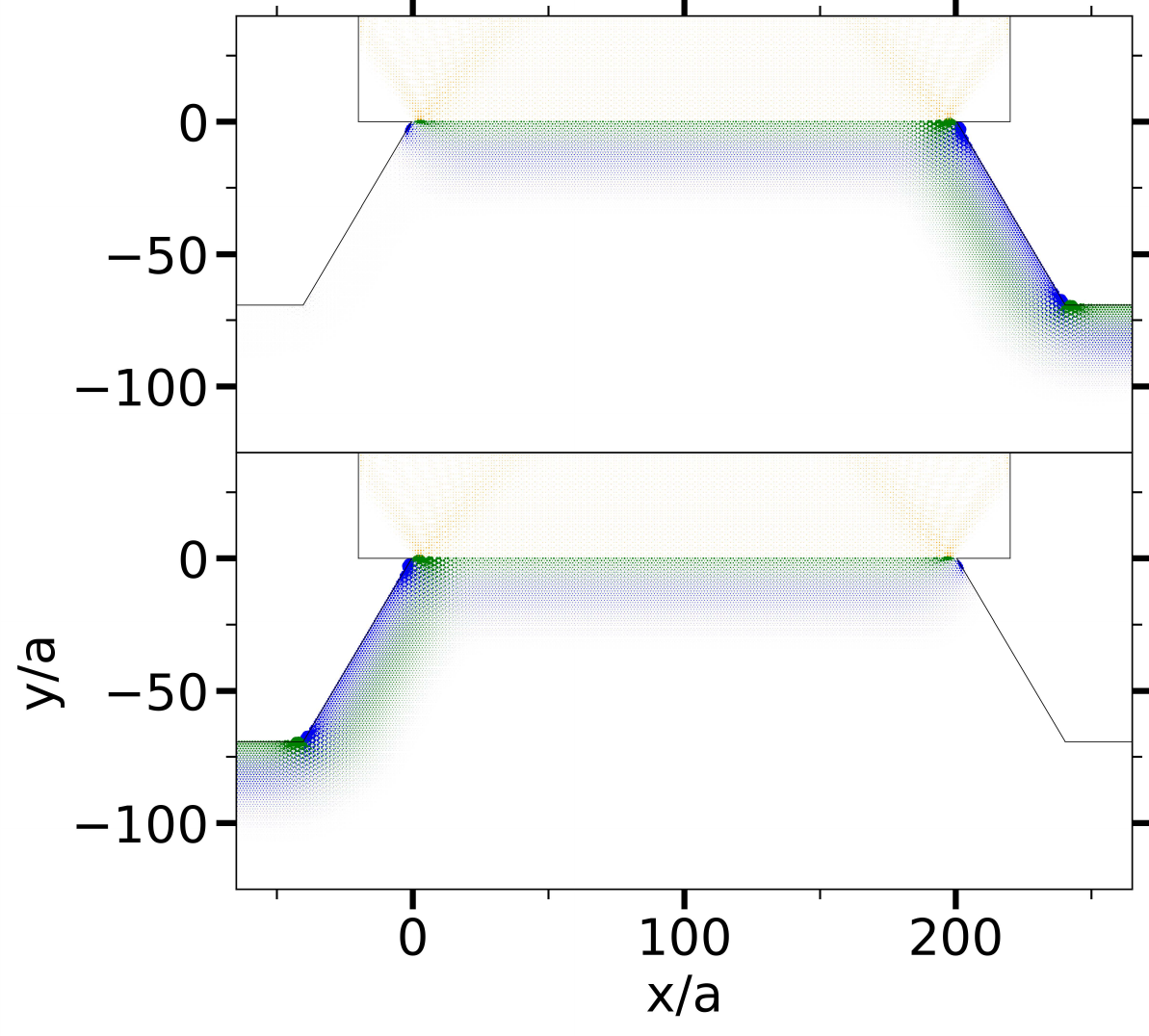}
    \caption{Scattering state, $|\psi^{\beta}(x,y)|^2$ with $\beta\!=\!e,h$,  for a \emph{sparse-stitched} zigzag trapezoid nanostructure at (a)~full or (b)~partial transparency.  In each panel the electron (hole) component is in the upper (lower) part \cite{PlottingMethod}. 
At the full-transparency interface, intervalley scattering is absent. At partial transparency, intervalley scattering appears at both interface corners, leading to interference effects. 
[$t_{NS}/t\!=\!0.3$ and $\mu_\textrm{gr}$ is for the peak near $\mu_\textrm{gr}/E_L=0.5$ in Fig.\,\ref{fig:Transport-sparse}(b) (green line); parameters otherwise standard (Sec.\,\ref{subsec:Parameters}).] 
    }
    \label{fig:SparsePsi1}
\end{figure}

\stepcounter{Sfigure}\begin{figure*}    
\includegraphics[width=3.4in]{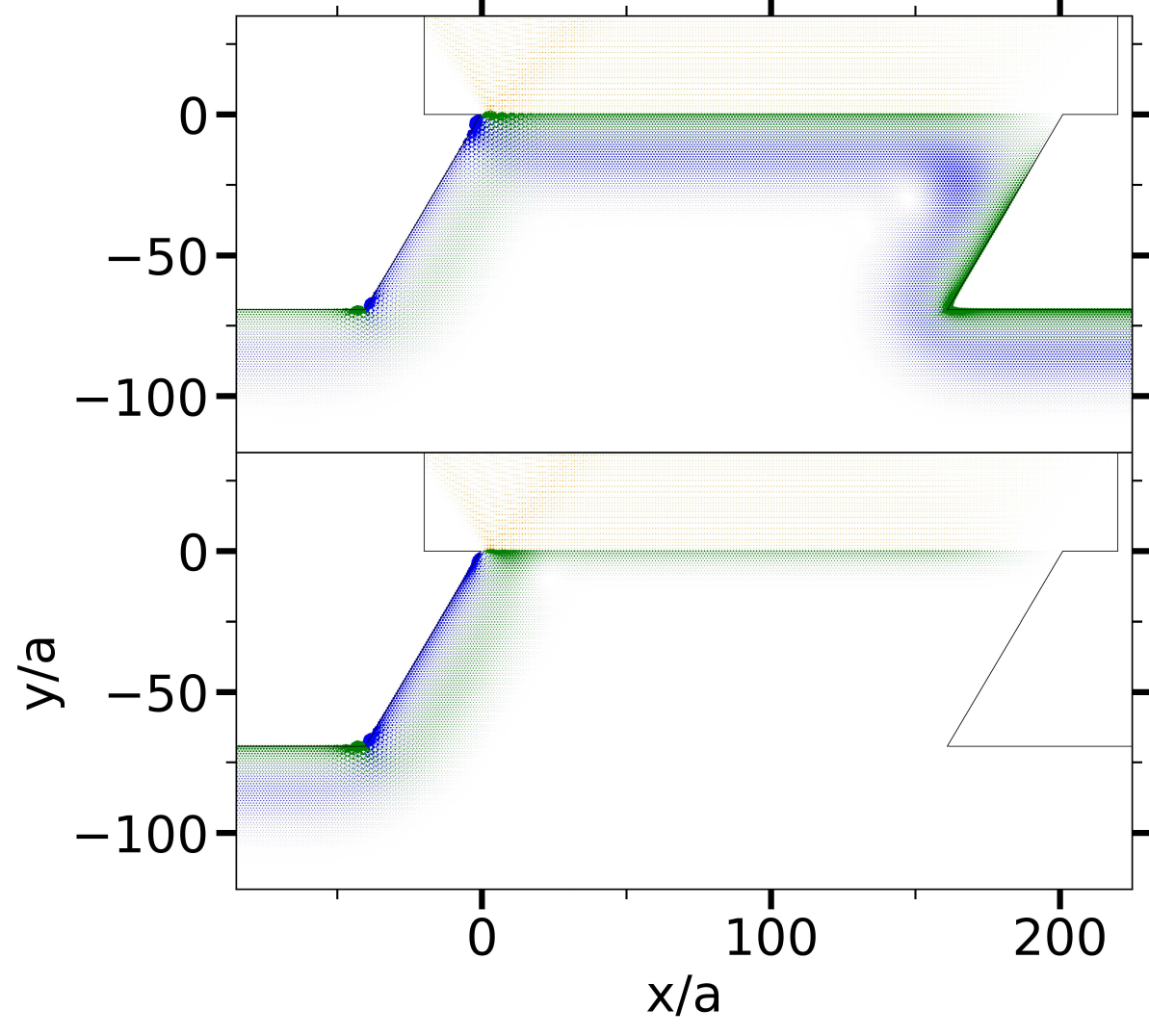}\hfill
\includegraphics[width=3.4in]{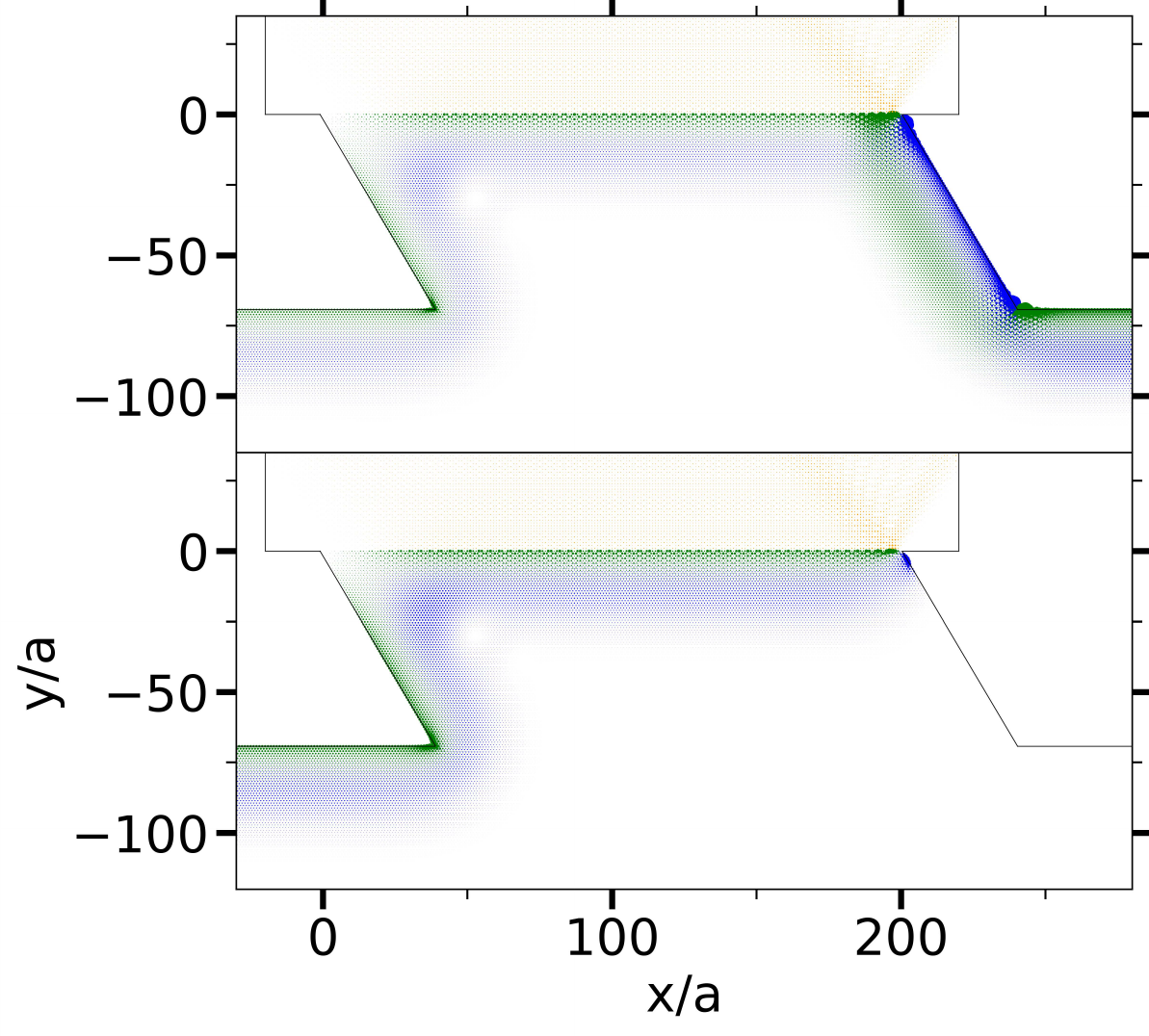}\\
\includegraphics[width=3.4in]{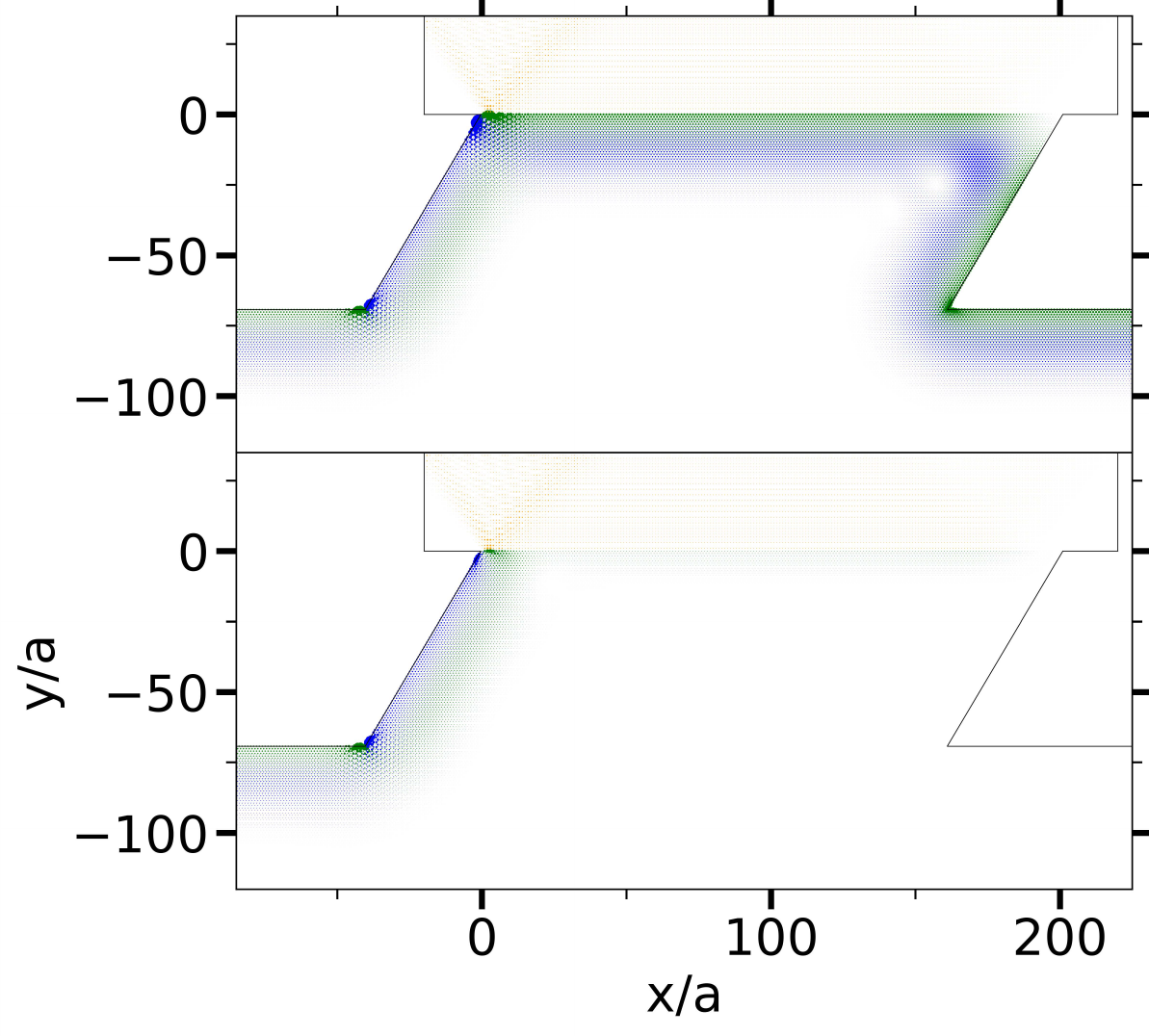}\hfill
\includegraphics[width=3.4in]{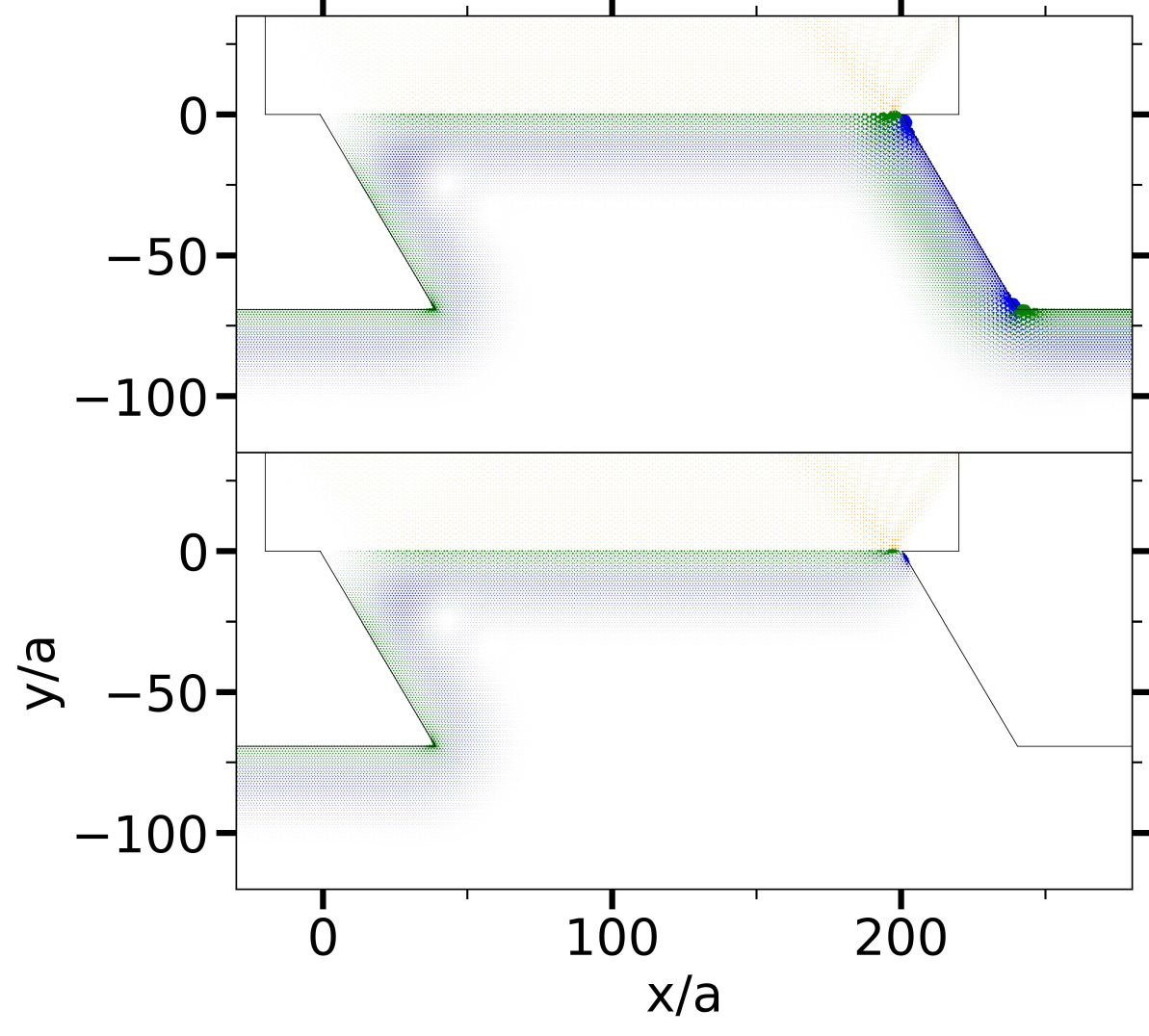}
\caption{Scattering state, $|\psi^{\beta}(x,y)|^2$ with $\beta\!=\!e,h$, for a \emph{sparse-stitched} zigzag parallelogram at reduced interface transparency ($t_{NS}/t\!=\!0.3$) and 
(row 1) low filling, $\mu_\textrm{gr}/E_L\!=\!1/4$, or (row 2) mid-Landau-gap, $\mu_\textrm{gr}/E_L\!=\!1/2$. 
In each panel the electron (hole) component is in the upper (lower) part \cite{PlottingMethod}. 
Despite weak electron-hole hybridization along the interface, the hole component dominates in the output, produced by Andreev intervalley scattering.  For our canonically oriented parallelogram (first column), Andreev intervalley scattering occurs at the downstream corner of the interface (left-hand side).  For the parallelogram in reverse orientation (second column), the intervalley-scattering corner is now upstream of the interface (right-hand side) and both interface modes are populated (note beating).  The final output intensity is, of course, independent of the parallelogram's orientation. 
(Parameters otherwise standard, see Sec.\,\ref{subsec:Parameters}.)  
}
    \label{fig:SparsePsi2}
\end{figure*}

\begin{figure}
\includegraphics[width=3.2in]{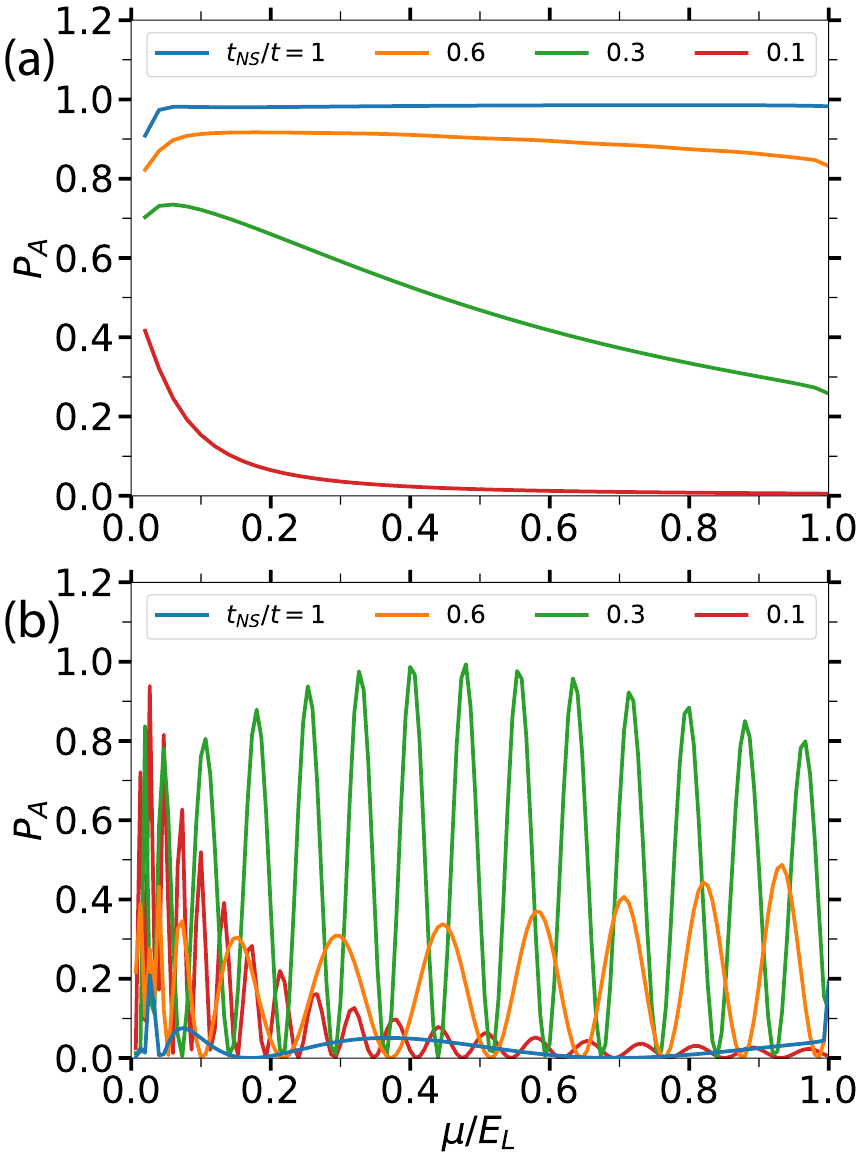}
\caption{Andreev conversion probability, $P_{he}$, as a function of the graphene doping $\mu_\textrm{gr}/E_{L}$ for two \emph{sparse-stitched} zigzag nanostructures: (a) parallelogram and (b) trapezoid.  
The strength of graphene-superconductor coupling, $t_{NS}/t$, is indicated by the line color.  Oscillations caused by interference are present when the corners at the entrance and exit of the interface \emph{both} cause intervalley scattering (trapezoid).  
At full interface coupling, $P_{he}(\mu_\textrm{gr})$ is typically constant, either $0$ or $1$.  At partial coupling, intervalley-scattering occurs at some corners, leading to rapid oscillations in the trapezoid.  Compare to dense-stitching results in Fig.\,\ref{fig:SummaryPA}.    
(Parameters otherwise standard, see Sec.\,\ref{subsec:Parameters}.)
}
\label{fig:Transport-sparse}
\end{figure}

Scattering states are shown in Figs.\,\ref{fig:SparsePsi1} and \ref{fig:SparsePsi2} for the all-zigzag trapezoid and parallelogram nanostructures, respectively.  For the trapezoid, as in the dense-stitched case (Fig.\,\ref{fig:Trap-withS}), the electron component connects smoothly in to and out of the interface when the graphene and superconductor are well-coupled via a transparent interface, implying the absence of any Andreev conversion in agreement with Eq.\,(\ref{eq:Phe-ideal-AkhBeen}).  
When the interface becomes partially transparent, intervalley scattering occurs at both ends of the interface, leading to beating of the two $e$-$h$ hybrid modes and oscillations in $P_{he}$ (see Fig.\,\ref{fig:Transport-sparse}).

Four scattering states are shown for the sparse-stitched parallelogram in 
Fig.\,\ref{fig:SparsePsi2}, two for the usual orientation of the parallelogram (first column) and two for the reversed orientation (second column), analogous to the dense-stitched results in Figs.\,\ref{fig:Paralel1-reduced} and \ref{fig:Paralel2-reduced}, respectively.  
The most striking result is that in the first column there is substantial Andreev conversion even though the $e$-$h$ hybridization at the interface is small.  Andreev conversion in these two cases is produced by Andreev intervalley scattering at the downstream corner of the interface.  

The wavefunction in the reversed parallelogram is quite different (Fig.\,\ref{fig:SparsePsi2} second column): here intervalley scattering at the upstream corner allows connection to the mostly hole mode along the interface, thus both modes are populated.  Because the weight in the electron and hole components is comparable, it appears at first glance that there is good hybridization along the interface.  However, closer examination reveals that this results from comparable occupation of the $e$-like mode and $h$-like mode along the interface; indeed, beating between the two modes can be seen.  Thus Andreev intervalley scattering is present in both orientations of the parallelogram: in the usual orientation it leads to a hole component on the outgoing QH edge even though the hybridization along the interface is poor, and in the reverse orientation it leads to comparable population of the two (weakly hybridized) interface modes. 

\stepcounter{Sfigure}

The Andreev conversion probability for sparse-stitched zigzag parallelogram and trapezoid structures are shown in Fig.\,\ref{fig:Transport-sparse}.  (For the parallelogram, $\theta_1\!=\!\theta_2\!=\!60^\circ$, while for the trapezoid $\theta_1\!=\!120$ and $\theta_2\!=\!60^\circ$.)  When the interface is partially transparent, $P_{he}(\mu_\textrm{gr})$ is smooth for the structure with only one intervalley-scattering corner (parallelogram) but oscillates rapidly when both corners scatter (trapezoid). 

\begin{figure}
\includegraphics[width=3.2in]{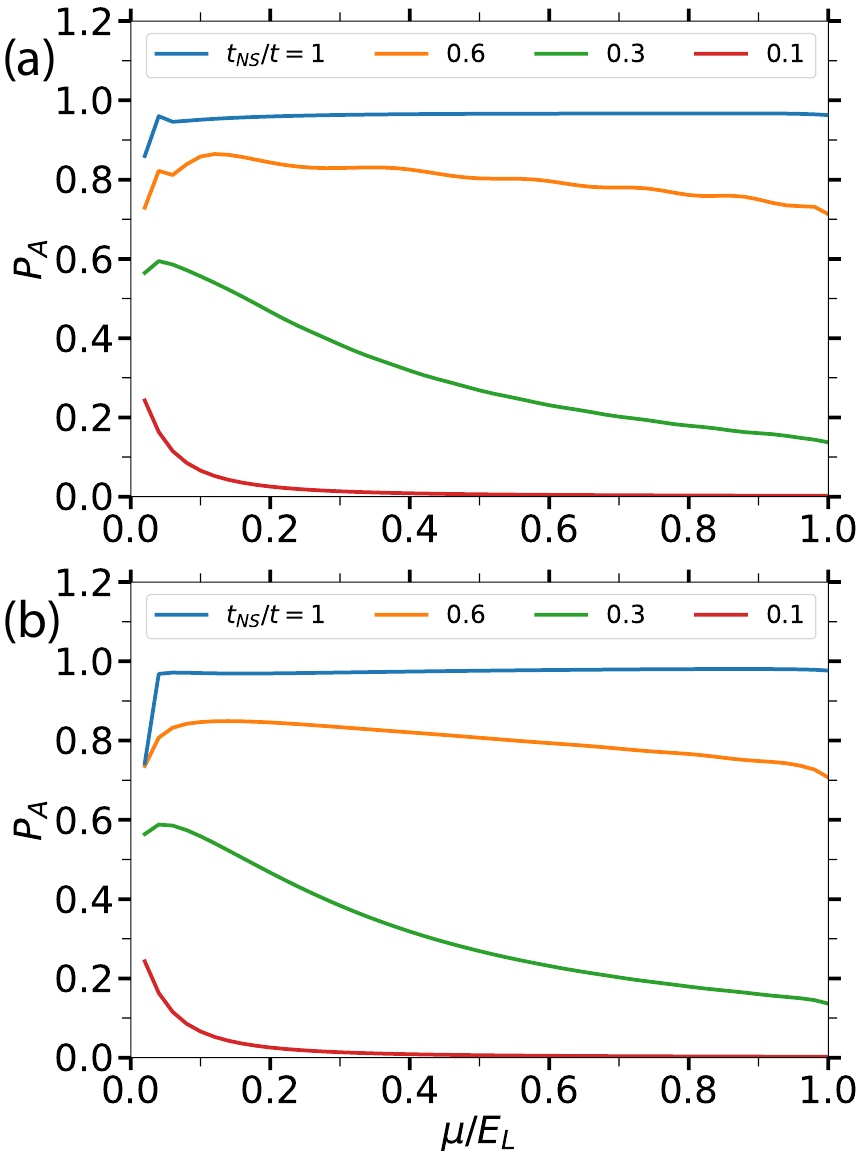}
\caption{Andreev conversion probability, $P_{he}$, as a function of the graphene doping $\mu_\textrm{gr}/E_{L}$ for two \emph{sparse-stitched} nanostructures that incorporate non-zigzag edges: (a) a $\theta_1\!=\!\theta_2\!=\!75^\circ$ parallelogram and (b) $\theta_1\!=\!45 ^\circ$, $\theta_2\!=\!75 ^\circ$.  The strength of graphene-superconductor coupling, $t_{NS}/t$, is indicated by the line color.  For full-transparency ($t_{NS}/t\!=\!1$), the Andreev conversion is nearly perfect for both structures, independent of doping.  
%Results are qualitatively similar to those for the zigzag parallelogram in Fig.\,\ref{fig:Transport-sparse}(a).  
Compare to dense-stitching results in Fig.\,\ref{fig:ArbAnglePhe}  (Standard parameters, see Sec.\,\ref{subsec:Parameters}.)
}
\label{fig:ArbAnglePhe-sparse}
\end{figure}

Fig.\,\ref{fig:ArbAnglePhe-sparse} shows the corresponding $P_{he}(\mu_\textrm{gr})$ for two sparse-stiched nanostructures that incorporate \emph{non-zigzag} edges.  The magnitude of the Andreev conversions tends to be smaller for non-zigzag structures, but the results are very similar qualitatively. 

\begin{figure}
\includegraphics[width=3.5in]{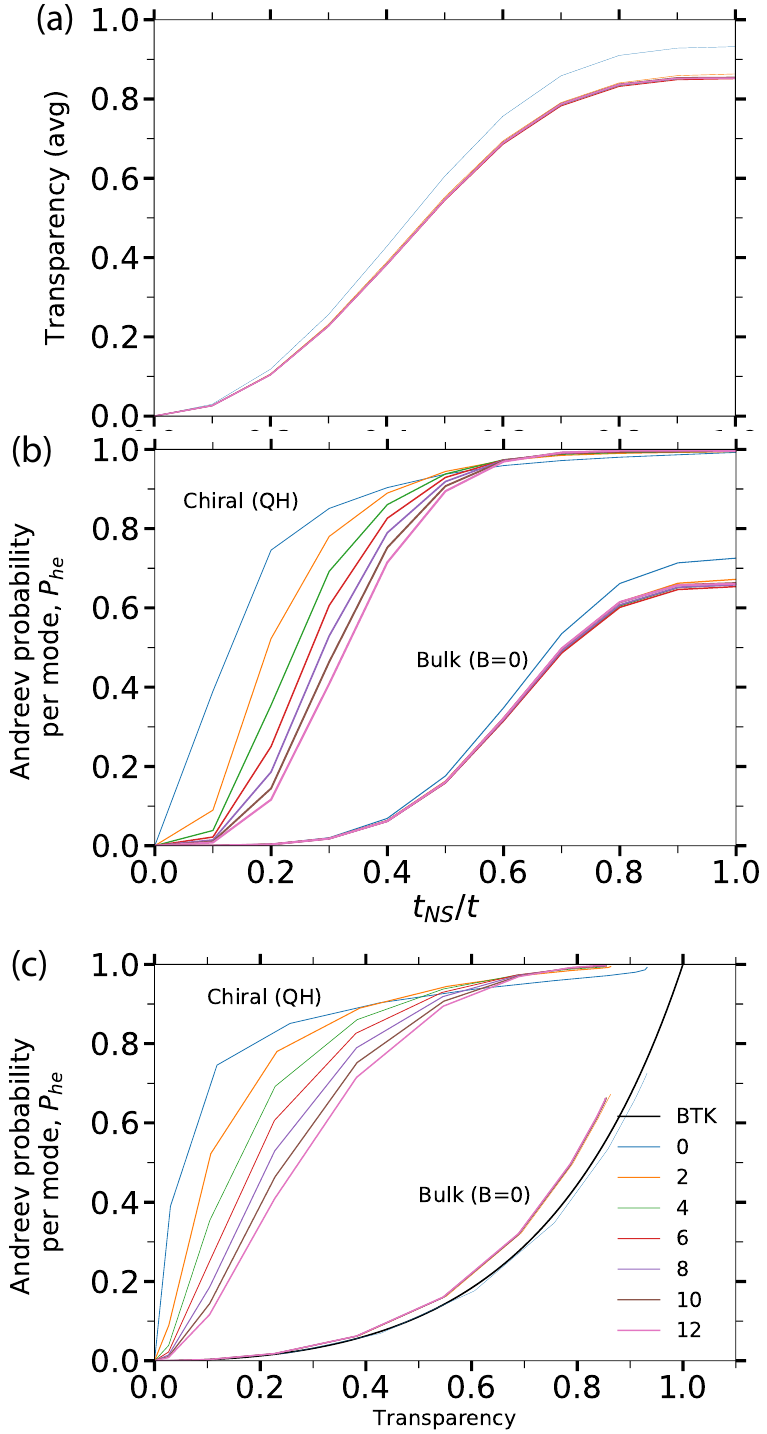}
\caption{Dependence of Andreev conversion on interface transparency, both at $B\!=\!0$ and in the QH regime, for several values of graphene doping (color-coded lines).  A zigzag parallelogram with a \emph{dense-stitched} interface is used.   
(a) Transparency---the two-terminal transmission probability through the interface with $B\!=\!\Delta\!=\!0$---as a function of the graphene-superconductor coupling, $t_{NS}/t$.  (b) Andreev probability per mode as graphene-superconductor coupling is varied.  (c) Andreev probability per mode reexpressed through the transparency from (a); the black line is the BTK result \citep{BTK-PRB82} for comparison.  The colors label graphene chemical potentials that uniformly sample 5--86\% of the LLL.  The zero field results are computed in a two-terminal nanoribbon geometry and averaged over several lengths of the interface.  Compare to sparse-stitched results in Fig.\,\ref{fig:transparency}.
}
\label{fig:transparency-dense}
\end{figure}

In the main text, Andreev conversion as a function of transparency of the interface is presented for a sparse-stitched interface (Fig.\,\ref{fig:transparency}).  Here, Fig.\,\ref{fig:transparency-dense} shows the corresponding analysis for a dense-stitched interface. 
There is again remarkable agreement between our tight-binding calculations at $B\!=\!0$ and the simple BTK result \citep{BTK-PRB82} using a 1D model with parabolic dispersion for both materials.  Andreev conversion is clearly more robust in the QH regime, especially for low doping, and shows some variation with $\mu_\textrm{gr}$.

\stepcounter{Sfigure} 

\stepcounter{Sfigure}

\end{document}